\documentclass[final,hidelinks,onefignum,onetabnum,final]{siamart250211}

\graphicspath{{figures/}}

\newcommand{\Tbind}{T_{*}}
\newcommand{\Pbind}{P_{*}}
\newcommand{\Jbind}{J_{*}}
\newcommand{\Pn}{P_{*, {\rm n}}}
\newcommand{\Pc}{P_{*, {\rm c}}}
\newcommand{\taun}{\tau_{*, {\rm n}}}
\newcommand{\tauc}{\tau_{*, {\rm c}}}
\newcommand{\x}{\mathbf{x}}
\newcommand{\taubind}{\tau_{*}}
\newcommand{\sigmabind}{\sigma_{*}}
\newcommand{\CVbind}{\text{CV}^{*}}
\newcommand{\Fc}{F_{\rm c}}
\def\dd{\textrm{d}}
\usepackage{accents}
\usepackage{tikz-cd}
\usepackage{chemfig}

\renewcommand{\v}[1]{{\mathbf{#1}}}

\newcommand{\mc}[1]{{\mathcal{#1}}}
\newcommand{\msf}[1]{{\mathsf{#1}}}

\definecolor{jade}{RGB}{0, 168, 107}

\usepackage{lipsum}
\usepackage{amsfonts}
\usepackage{graphicx}
\usepackage{epstopdf}
\usepackage{algorithmic}
\usepackage{chemfig}
\usepackage{chemformula}
\usepackage{caption}
\usepackage{grffile}
\usepackage{hyperref}
\usepackage{mathtools}
\usepackage{chemmacros}
\chemsetup{modules={reactions}}
\usepackage{chemfig}
\usepackage{tikz}
\usetikzlibrary{positioning,calc,arrows.meta}
\usepackage{verbatim}
\usepackage{bbm}
\usepackage[labelfont=bf]{caption,subcaption}
\ifpdf
  \DeclareGraphicsExtensions{.eps,.pdf,.png,.jpg}
\else
  \DeclareGraphicsExtensions{.eps}
\fi
\usepackage{dsfont}
\usepackage{nicefrac}       
\usepackage{xfrac}
\usepackage{nccmath}
\usepackage[sort,compress]{cite}

\usepackage[draft]{changes}

\usepackage{comment}
\newsiamremark{remark}{Remark}
\newsiamremark{hypothesis}{Hypothesis}
\crefname{hypothesis}{Hypothesis}{Hypotheses}
\newsiamthm{claim}{Claim}

\headers{T cell search: diffusion, death and activation}{Tony Wong,
  Ikchang Cho, Maria R. D'Orsogna, and Tom Chou}

\title{First Passage Times to
 T cell activation}

\author{Tony Wong\thanks{Department of Mathematics, University of
    California, Los Angeles. (\email{tonywong@math.ucla.edu}).}  \and
  Ikchang Cho\thanks{The University of Chicago.
    (\email{ikchangcho@uchicago.edu}).}  \and Maria
  R. D'Orsogna\thanks{Department of Mathematics, California State
    University at Northridge (\email{dorsogna@csun.edu}). } \and Tom
  Chou \thanks{Department of Computational Medicine, University of
    California, Los Angeles.}(\email{tomchou@edu.ucla}). }

\usepackage{amsopn}

\ifpdf
\hypersetup{
  pdftitle={None},
  pdfauthor={Tony Wong, Ikchang Cho, Maria R. D'Orsogna, and Tom Chou}
}
\fi

\begin{document}
\maketitle

\begin{abstract}
Effective recognition of foreign antigens by the adaptive immune
system relies on T cells being activated by antigen-presenting cells
(APCs) in lymph nodes.  Here, diffusing T cells may encounter cognate
APCs that present matching antigen fragments or non-cognate ones that
do not; they are also subject to degradation.  We develop a stochastic
model in which T cell-APCs interact via a sequence of recognition
steps, represented as a multistage Markov chain.  T cells are
successfully activated only if the terminal state associated with a
cognate APC is reached.  We compute the probability of successful
activation in the presence of interfering non-cognate APCs, T cell
degradation, and lymph node exit, and analyze the mean first-passage
time to activation. We also incorporate a kinetic proofreading
mechanism that enables state resetting, and show how this enhances
specificity toward cognate APCs.

\textbf{Relevance to Life Sciences} We present a quantitative
framework to study T cell activation within the lymph node that
integrates diffusion, the presence and abundance of cognate and
non-cognate APCs, and T cell death and exit from the lymph
node. Relevant spatio-temporal parameters, such as T cell diffusivity
and residence time within the lymph node, are estimated from existing
literature.  Quantification of the activation probability and time to
first activation provide fundamental insights into the onset of the
adaptive immune response.

\textbf{Mathematical Content} T cell recognition is modeled as a
multistage Markov process, coupled with spatial diffusion, exit, and
death. All four processes are represented through a system of partial
differential equations that are analyzed under Robin and Neumann
boundary conditions.  Using first-passage time theory, we calculate
activation probabilities and mean activation times.  We also show how
kinetic proofreading through stochastic resetting enhances
specificity.

\end{abstract}



\section{Introduction and Background}
\label{sec:intro}

The adaptive immune system plays a central role in defending an
organism from disease.  Key components are antigen-presenting cells
(APCs) and T cells that co-localize in lymph nodes where they interact
to trigger proliferation or B cell signaling \cite{Parkin2001}.  APCs
capture antigens, short amino-acid sequences that are part of larger
proteins, from foreign agents encountered throughout the body.
Enzymatic degradation turns these antigens into smaller peptides that
are presented to the major histocompatibility complexes (MHCs) on the
APC surfaces for T cells to recognize.

T cells constitute a highly diversified population. Each of them
expresses a specific surface receptor (TCR) that can only recognize a
small subset of ``cognate'' antigens.  Successful recognition involves
several biochemical interactions between TCRs and the antigen-loaded
MHCs that include conformational changes which in turn trigger
downstream signaling events.  Upon recognition, na\"ive T cells become
activated and initiate an immune response by proliferating and
differentiating into effector T cells such as cytotoxic T cells that
eliminate pathogens, and helper T cells that activate other immune
cells; most of them migrate from the lymph node to peripheral tissues
to kill the infected host cells.  Some activated T cells become
long-lived memory cells that help trigger an effective and rapid
immune response if the same pathogen is encountered again.  Each T
cell is associated to a small number of cognate APCs (cAPCs), all the
others are not recognized and are referred to as non-cognate APCs
(nAPCs). It is estimated that a given APC is cognate to one in
$10^5–10^6$ T cells
\cite{Blattman2002,Jenkins2010,Yu2015,krummel_gerard_2016}; the
inverse of this quantity is known as the precursor frequency. Given
how rare it is for a T cell to encounter its cAPC, the problem is
sometimes known as that of ``searching for a needle in a haystack''
\cite{miller_parker_2003}.

Most T cells are produced in the bone marrow, mature in the thymus,
and are transported via the bloodstream to the approximately 600 lymph
nodes in the human body. T cells are in continuous recirculation: from
the blood they reach a lymph node and search for their cAPC. If the
search is not successful within 12 to 24 hours, they return to the
bloodstream, enter a different lymph node and repeat the cycle
\cite{Katakai2013,Miyasaka2016,Ugur2019,grigorova_cyster_2010}.  The
lifespan of na\"ive, unactivated T cells ranges from weeks to years,
depending on age and health status \cite{Vrisekoop2008, Borghans2017}.
On the contrary, once a foreign antigen has been acquired, an APC will
migrate from the exposed tissue (such as the skin) to the closest
lymph node through lymphatic vessels and remain there
\cite{Randolph2005}. The typical lifespan of an APC within a lymph
node is two to six days \cite{Tomura2014, Kitano2016}.

Recent advances in 3D imaging, flow cytometry and quantitative PCR
have been used to shed light on how T cells interact with APCs, particularly dendritic cells \cite{Groom2015, Ozulumba2023}.  Most T
cells and APCs co-localize in the ``T cell zone,'' a specialized
sub-compartment that occupies a large portion of the lymph node \cite{Banchereau1998}.  T cells within this compartment do not follow
chemotactic gradients but instead encounter antigen-presenting cells
(APCs) through independent, random motion. Their movement is guided by
an underlying network of fibroblastic reticular cells, which provides
structural support to the lymph node and helps organize cell
trafficking within it \cite{miller_parker_2003, Bajenoff2006,
  Katakai2004, Link2007}.
The speed, persistence times, and turning angles of T cells have been
quantified, revealing that within the T cell zone, T cells are much
more motile than dendritic and B cells \cite{miller_cahalan_2002,
  Torres2023}.  In addition, dendrites emanating from the cellular core of
dendritic cells are highly dynamic, increasing their effective spatial
extent.  As a result, the contact frequency between T cells and APCs
is elevated, leading to an efficient scanning process
\cite{Westermann2005}.  For example, a dendritic cell can engage with
up to 80 T cells per minute with the typical contact between a T cell
and a nAPC lasting roughly 3 minutes before dissociation
\cite{miller_parker_2004}.  Upon encountering a cAPC, however, a T
cell will arrest its motion to allow for biochemical and
conformational changes that stabilize the low-affinity TCR-MHC contact
to take place; this association can last more than 15 hours before the
immune response is triggered \cite{stoll_germain_2002}.

Mathematically, the movement of T cells in lymph nodes has been
studied via computational models based on two-photon microscopy
imaging \cite{Molina2021, Beltman_2007, beauchemin_perelson_2007,
  donovan_lythe_2012, fricke_cannon_2016}.  For example, the residence
time of a T cell interacting with a nAPCs has been fitted to an
exponential distribution whose mean depends on the specific T cell
type \cite{mandl_ronald_2012}.  Other theoretical studies propose
different types of random walks, such as Brownian motion
\cite{celli_bousso_2012, delgado_coombs_2015}, L\'evy walks
\cite{harris_hunter_2012}, and velocity-jump models
\cite{preston_pritchard_2006}.  Typically, cAPCs are modeled as a
finite set of small, stationary, and well-separated target sites;
encounters are defined as a T cell reaching or coming within a given
distance from these sites \cite{celli_bousso_2012}.  In these studies,
the presence of nAPCs is often neglected \cite{mckeithan_1995,
  davis_anton_2006, lever_dushek_2014}.  The probability of a T cell
encountering its cAPC within a fixed time has also been used to
estimate the likelihood of initiating the adaptive immune response
\cite{preston_pritchard_2006, celli_bousso_2012}.

\section{Mathematical model and analysis}

\begin{figure}[t]
\centering
\includegraphics[width=0.93\textwidth]{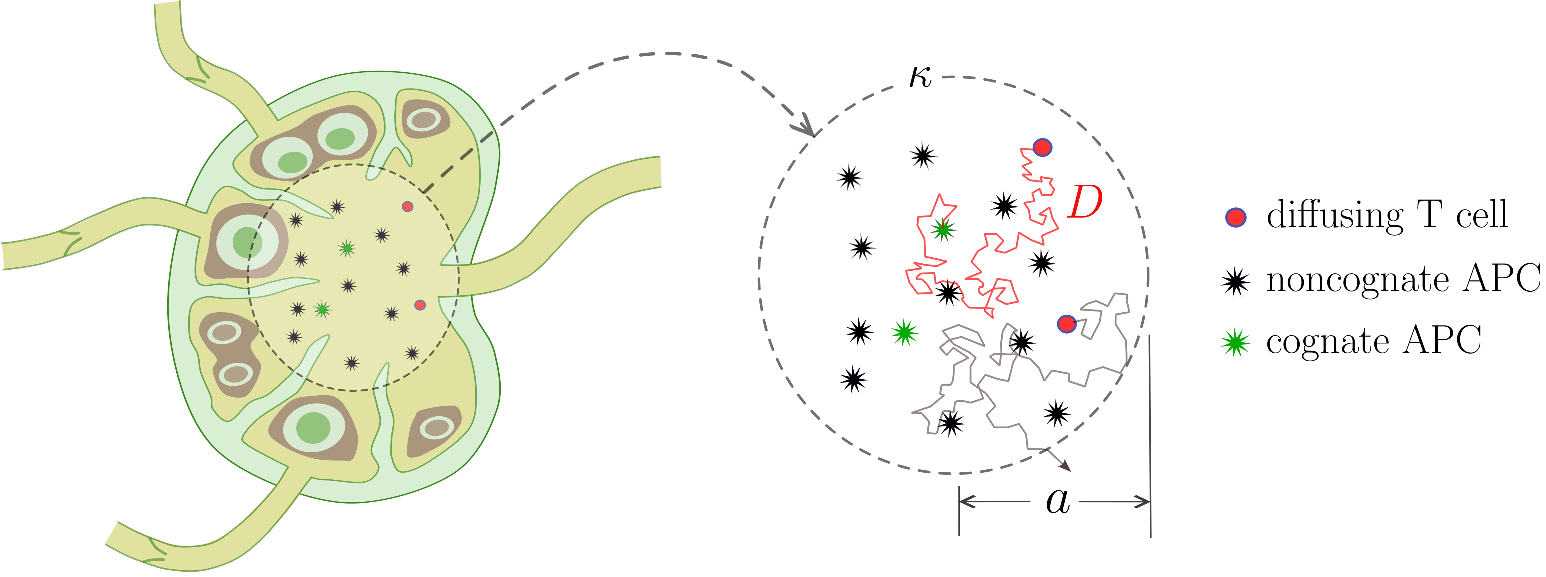}
\caption{\textbf{Spatial representation of our model.}  A schematic of
  a lymph node is shown on the left. The T cell zone, where T cells
  (red dots) diffuse and encounter cognate (green stars) or noncognate
  (black stars) APCs, is approximated as a sphere of radius $a$ in the
  schematic on the right. The diffusion constant $D$ is assumed to be
  uniform.  Removal of T cells occurs via two mechanisms: exit through
  the boundary of the T cell zone, and death or egress from the
  interior. The latter process is facilitated by specialized lymphatic
  channels not shown.}
    \label{LYMPH_NODE}
\end{figure}

We develop a mathematical framework to study antigen induced T cell
activation where the presence of both cognate and non-cognate APCs are
explicitly included. Other features are T cell death and egress from
lymphatic channels, the relative scarcity of cAPCs compared to nAPCs,
and non-trivial activation mechanisms such as multi-stage binding or
kinetic proof-reading.  We quantify the statistics of the conditional
cAPC-induced activation times of T cells \cite{delgado_coombs_2015} in
the dominating presence of nAPCs.  T cells are assumed to be point
particles that diffuse in a three-dimensional, spherical lymph node
compartment, uniformly populated by APCs.  Quantities of interest
include the probability that a T cell is activated by its cAPC before exiting the
T cell zone, and the conditional first passage time to full activation
\cite{redner_2001, dorsogna2005, chou_dorsogna_2014,
  iyer_zilman_2016}. The overall geometry of our model is shown in
Fig.~\ref{LYMPH_NODE}, where for simplicity the T cell zone is modeled
as a sphere.

In \cref{multi-stage} we present a reversible, multi-stage, two-arm
model, in which a T cell diffusing within a sphere can bind to either
a non-cognate APC (nAPC) or a cognate APC (cAPC).  The initial contact
state is followed by a multi-state recognition process that terminates
at state $N$. Transitions between states are reversible, with the
exception of the final state at the end of the cAPC arm; here the T
cell is immediately activated and can no longer transition back.  We
evaluate the overall activation probability and define the conditional
moments of the activation time. From the latter, we calculate the mean
and variance of the first time for a T cell to be activated by a cAPC
under both Neumann and Robin boundary conditions at the spherical
boundary representing the T cell zone.  In \cref{sec:KP} we consider
an alternative scenario where a T cell can fully bind to both nAPCs
and cAPCs once their respective multi-stage chains have been
traversed. However, intermediate states in each chain can ``reset'',
returning the T cell to its initial state of engagement with the nAPC
or cAPC. This recycling represents a kinetic proofreading mechanism
that can increase sensitivity to kinetic parameters, allowing for
higher specificity.

\section{Reversible multi-stage two-arm model} 
\label{multi-stage}

We assume recognition involves T cells engaging with APCs through
multiple interaction steps that lead to full activation only in the
case of cAPCs.  Similar models have been used to study viral entry
into cells \cite{Dorsogna2007,Gibbons2010}.  A schematic of the model
with $N$ intermediate states between a T cell and either APC is shown
in \eqref{eqn:transmission_pathway_APC}.  Here, the free T cell,
denoted by $\mathsf{T}_0$, can bind to a nAPC at rate $K_\mathrm{n}$
to form the first bound state $\mathsf{N}_1$, or with a cAPC at rate
$f K_\mathrm{n}$ to form the first bound state $\mathsf{C}_1$.
Microscopically, $K_{\rm n} = K_{0,1} {\cal N}$ and $K_{\rm c} =
K_{0,1} {\cal C}$, where $K_{0,1}$ is an intrinsic attachment
prefactor, and $\cal N$ and $\cal C$ denote the concentrations of
available nAPCs and cAPCs, respectively. Rewriting $K_{\rm c} = ({\cal
  {C}} / {\cal N}) K_{\rm n} \equiv f K_{\rm {n}}$, the parameter $f$
can be interpreted as the ratio of the two concentrations; since the
concentration of nAPCs is much greater than that of cAPCs, $f \ll
1$. The unbinding rates are $K_{1,0}$ in both arms.

\begin{center}
\tikzset{
   forward arrow/.style={
   -{Stealth[harpoon, length=4pt]},
    draw, 
     line width=0.4pt 
  },
  species/.style={
    font=\normalsize
  },
}

\begin{equation}
  \begin{tikzpicture}[baseline=(current  bounding  box.center),
  species/.style={font=\normalsize, inner sep=2pt, outer sep=2pt},
  small/.style={font=\scriptsize},
  fwd/.style={-{Stealth[harpoon,length=4pt]}, line width=0.45pt},
  bwd/.style={-{Stealth[harpoon,length=4pt]}, line width=0.45pt}
]
    \label{eqn:transmission_pathway_APC}

\def\theta{40}          
\def\r{1.8cm}           
\def\step{1.5cm}        
\def\shortenH{0.15}     
\def\offset{1.6pt}      

\node[species] (T0) at (0,0) {$\mathsf{T}_0$};

\path (T0) --++(\theta:\r)  node[species] (N1) {$\mathsf{N}_1$};
\path (T0) --++(-\theta:\r) node[species] (C1) {$\mathsf{C}_1$};

\node[species,right=\step of N1] (N2) {$\mathsf{N}_2$};
\node[species,right=\step of N2] (N3) {$\cdots$};
\node[species,right=\step of N3] (NNm1) {$\mathsf{N}_{N-1}$}; 
\node[species,right=\step of NNm1] (N4) {$\mathsf{N}_N$};

\node[species,right=\step of C1] (C2) {$\mathsf{C}_2$};
\node[species,right=\step of C2] (C3) {$\cdots$};
\node[species,right=\step of C3] (CCm1) {$\mathsf{C}_{N-1}$}; 
\node[species,right=\step of CCm1] (C4) {$\mathsf{C}_N$};

\draw[fwd, shorten >=2pt, shorten <=2pt, transform canvas={yshift=+\offset}] 
      (T0) -- (N1) node[midway,sloped,above,small,pos=0.5] {$K_{\mathrm{n}}$};
\draw[bwd, shorten >=2pt, shorten <=2pt, transform canvas={yshift=-\offset}] 
      (N1) -- (T0) node[midway,sloped,below,small,pos=0.5] {$K_{1,0}$};

\draw[fwd, shorten >=2pt, shorten <=2pt, transform canvas={yshift=+\offset}] 
      (T0) -- (C1) node[midway,sloped,above,small,pos=0.5] {$fK_{\mathrm{n}}$};
\draw[bwd, shorten >=2pt, shorten <=2pt, transform canvas={yshift=-\offset}] 
      (C1) -- (T0) node[midway,sloped,below,small,pos=0.5] {$K_{1,0}$};

\foreach \A/\B in {N1/N2,N2/N3,N3/NNm1,NNm1/N4}{
  \draw[bwd, shorten >=\step*\shortenH, shorten <=\step*\shortenH, transform canvas={yshift=-\offset}] 
        (\B.west) -- (\A.east) node[midway,below,small] {$Q$};
  \draw[fwd, shorten >=\step*\shortenH, shorten <=\step*\shortenH, transform canvas={yshift=+\offset}] 
        (\A.east) -- (\B.west) node[midway,above,small] {$P$};
}

\foreach \i/\A/\B in {1/C1/C2,2/C2/C3,3/C3/CCm1,4/CCm1/C4}{
  \ifnum\i<4
    \draw[bwd, shorten >=\step*\shortenH, shorten <=\step*\shortenH, transform canvas={yshift=-\offset}] 
          (\B.west) -- (\A.east) node[midway,below,small] {$Q$};
  \fi
  \draw[fwd, shorten >=\step*\shortenH, shorten <=\step*\shortenH, transform canvas={yshift=+\offset}] 
        (\A.east) -- (\B.west) node[midway,above,small] {$P$};
}

\end{tikzpicture}
\end{equation}
\end{center}

The nAPC interaction chain, shown in the upper arm of the scheme in
\eqref{eqn:transmission_pathway_APC} contains a sequence of
intermediate bound states, $\mathsf{N}_2,\dots\mathsf{N}_{N-1}$, and
is fully reversible, with the bidirectional forward and backward rates
given by $P$ and $Q$, respectively \cite{Hurtado2021}.  The reflecting
boundary condition at the last $\mathsf{N}_N$ state represents a
``dead-end'' of the interaction between a T cell and the nAPC, where
no further processing is triggered. The cAPC interaction chain is
modeled through a similar pathway with $N$ steps as shown in the lower
arm of \eqref{eqn:transmission_pathway_APC}.  Here, the absorbing
state $\mathsf{C}_N$ represents a T cell that is activated by its
cAPC.  For simplicity, we assume uniform binding and unbinding rates
$P,Q$ and compute the probability of activation and the activation
time statistics.

\subsection{Diffusion-kinetic equations and non-dimensionalization}
\label{sec:equations}

We now quantify the dynamics of the system, including motion in the T
cell zone and the kinetics depicted in
\eqref{eqn:transmission_pathway_APC}.  Since searcher T cells have no
prior information on the location of APCs within the lymph node
\cite{preston_pritchard_2006, donovan_lythe_2012}, we model them as
three-dimensional Brownian walkers and assume their motion is arrested
upon contact with an APC.  We introduce $\v{x}$ and $t$ as the spatial
and temporal variables, and denote the probability density
distribution of finding a T cell at location $\v{x}$ at time $t$ as
$\rho_0(\v{x},t)$. We further denote the probability density
distribution of a T cell bound to a nAPC or to a cAPC, respectively,
via the $N$-dimensional vectors $\v{n} (\v{x},t)$ and
$\v{c}(\v{x},t)$. The corresponding $n_i (\v{x},t)$ and $c_i
(\v{x},t)$ components represent the probability of having a T cell
bound to the nAPC or cAPC at the $i$-th state of engagement, with $1
\leq i \leq N$.  T cell activation is triggered at state $c_{\rm N}$.

Although the effective diffusion constant $D(\v{x})$ describing the
motion of the T cells may be spatially dependent, we impose it to be
uniform $D(\v{x}) = D$.  Similarly, we assume $K_{0,1}, K_{1,0}, \cal
N, \cal C$ are spatially homogeneous so that $K_{\rm n}$ and $f$ are
uniform as well.  Finally, we assume T cells exit from, or degrade
within the T cell zone at rate $\mu_0$.  The above assumptions yield
the following diffusion-kinetic equations

\begin{subequations}
\label{eqn:multi-stage}
\begin{eqnarray}
\partial_t \rho_0(\x,t) & = &  D \Delta \rho_0(\x,t)  - \left[\mu_0 + (1 + f) K_{0,1}\right]  \rho_0 
+  K_{1,0} (n_1 +  c_1), \label{eqn:multi-stage_PDE_rho0}\\[5pt]
 \partial_t \, \v{n}(\x,t) &=& \v{M}_\mathrm{n} \v{n} + K_{0,1} \rho_0 \, \v{e}_1,
 \label{eqn:multi-stage_PDE_nAPC} \\[5pt]
 \partial_t \, \v{c}(\x,t) & = & \v{M}_\mathrm{c} \v{c} + f K_{0,1} \rho_0,
 \v{e}_1. \label{eqn:multi-stage_PDE_cAPC}
\end{eqnarray}
\end{subequations}

\noindent
where $\v{e}_1 = (1,0,\dots,0)^T \in \mathbb{R}^N$ and
$\v{M}_\mathrm{n}$ and $\v{M}_\mathrm{c}$ are $N\times N$ tridiagonal
matrices consistent with the reaction schemes shown in
\eqref{eqn:transmission_pathway_APC}.  For simplicity, we describe
their elements in mathematical detail after first non-dimensionalizing
\eqref{eqn:multi-stage}. We model the T cell zone as a spherical
domain $\Omega$ of radius $a$ and impose Robin boundary conditions
\begin{equation}\label{eqn:robin_bc_dim}
D \, \v{n} \cdot  \nabla \rho_0 + 
K  \rho_0 = 0 \,, \quad \text{on} \quad \partial\Omega \,.
\end{equation}
The quantity $K > 0$ in \eqref{eqn:robin_bc_dim} represents the
convective velocity at the boundary.  The limit $K \to 0$ leads to the
perfectly reflecting, Neumann boundary condition $\v{n} \cdot \nabla
\rho_0 = 0$, where all traversing T cells are reflected to the
interior.  The opposite limit $K \to \infty$ leads to the perfectly
absorbing, Dirichlet boundary condition $\rho_0 = 0$ which represents
all T cells leaving the domain upon reaching its boundary.  The
partially absorbing, Robin boundary condition corresponds to a finite
$K$ and interpolates between the two limits.  Finally, we utilize the
initial condition
\begin{equation}\label{eqn:IC0}
\rho_0(\v{x},0) = \delta(\v{x}) \,, \quad \v{n} (\v{x},0) = 0 \,, \quad \v{c}(\v{x}, 0) = 0, \,
\end{equation}
so that at $t=0$ there are no bound APCs, and T cells are located at
the center of the lymph node.

To non-dimensionalize \eqref{eqn:multi-stage}--\eqref{eqn:IC0} we
define distances in terms of the radius of the T cell zone $a$ and
time in terms of the T cell -- nAPC detachment time $1/K_{1,0}$.  The
dimensionless variables are thus defined as
\begin{equation}
\begin{aligned}
\label{nondim} 
&\tilde{t} = K_{1,0} \, t \,, \, \quad 
\tilde{\v{x}} = \frac{\v{x}}{a}  \,, \quad  
\tilde{\mu}_0 =  \frac {\mu_0}{K_{1,0}}  \,, \quad  
p = \frac{P}{K_{1,0}}  \,, \quad  
q = \frac{Q}{K_{1,0}}  \,, \quad  
 \\[5pt]
& \tilde{D} = \frac{D}{a^2 K_{1,0}}, \,\,\,
k_{0,1} =  \frac{K_{\rm n}}{K_{1,0}}, 
 \,\,\, \tilde{K} = \frac{K}{a K_{1, 0}}, \,\,\,
\kappa = \frac{aK}{D}.
\end{aligned}
\end{equation}
Using estimates available from the literature, we set $a = 0.1$ cm,
$K_{1,0} = 1/3 \, \text{min}^{-1}$, $D = 60 \, \mu \mathrm{m}^2
\mathrm{min}^{-1}$, and $\mu_0 = 1/720 \, \text{min}^{-1}$. These
values correspond to $\tilde{\mu}_0 = 1/240$ and $\tilde{D} = 1.8
\times 10^{-4}$. Parameter estimates are discussed in detail in
Appendix \ref{parameters}. For notational simplicity, we henceforth
drop the tilde notation and find the dimensionless entries of
$\v{M}_\mathrm{n}$

%
\begin{equation}
  \label{eqn:multi-stage1}
  \begin{aligned}
  &\begin{aligned}
 [\v{M}_\mathrm{n}]_{i,i}  =
    \begin{cases}
      -(p+1) & \text{$i=1$} \\
      -(p+q)& \text{$2 \leq i \leq N-1$} \\
      -q & \text{$i=N$}
    \end{cases}
  \end{aligned}
  &\,\,\,  
  \begin{aligned}
 [\v{M}_\mathrm{n}]_{i,i-1} & = p \quad \enspace 2 \leq i \leq N, \, \\[5pt] [\v{M}_\mathrm{n}]_{i,i+1} & = q \quad \enspace 1 \leq i \leq N-1, 
  \end{aligned}
\end{aligned}
\end{equation}
while those for $\v{M}_\mathrm{c}$ are 
\begin{equation}
  \label{eqn:multi-stage2}
  \begin{aligned}
  & \begin{aligned}
 [\v{M}_\mathrm{c}]_{i,i}  =
    \begin{cases}
      -(p+1) & \text{$i=1$} \\
      -(p+q)& \text{$2 \leq i \leq N-1$} \\
      0& \text{$i=N$}
    \end{cases}
  \end{aligned}
  &\,\,\,  
  \begin{aligned}
&[\v{M}_\mathrm{c}]_{i,i-1} = p \quad \enspace 2 \leq i \leq N \\[5pt]
& [\v{M}_\mathrm{c}]_{i,i+1}  =
    \begin{cases}
      q & \text{$1 \leq i \leq N-2 $}\\
      0 & \text{$i=N-1$}.
    \end{cases}
  \end{aligned}
\end{aligned}
\end{equation}

%
$\v{M}_\mathrm{n}$ and $\v{M}_\mathrm{c}$ differ only in that the
last, activated state at the end of the cAPC chain is absorbing, while
the end-state of the nAPC chain is reflecting.  The full
non-dimensional model is defined in the three-dimensional ball of unit
radius $\Omega = B_1^3(0)$:

\begin{subequations}
\label{eqn:multi-stage_nd}
\begin{eqnarray}
\partial_t \rho_0(\x,t) & = &  D \Delta \rho_0(\x,t)  - \left[\mu_0 + (1 + f) k_{0,1}\right]  \rho_0 
+  n_1 +  c_1, \label{eqn:multi-stage_PDE_rho0_nd}\\[5pt]
 \partial_t \, \v{n}(\x,t) &=& \v{M}_\mathrm{n} \v{n} + k_{0,1} \rho_0 \, \v{e}_1,
 \label{eqn:multi-stage_PDE_nAPC_nd} \\[5pt]
 \partial_t \, \v{c}(\x,t) & = & \v{M}_\mathrm{c} \v{c} + f k_{0,1} \rho_0,
 \v{e}_1, \label{eqn:multi-stage_PDE_cAPC_nd} 
 \end{eqnarray}
\end{subequations}
 with boundary condition 
\begin{equation}
\label{eqn:robin_bc_dim_nd}
\v{n} \cdot  \nabla \rho_0 + \kappa  \rho_0 =  0 \,, \quad \text{on} \quad \partial\Omega \,.
\end{equation}
We now define the overall activation
flux $J_*(t)$ 

\begin{equation}\label{eqn:binding_flux_multi-stage}
  J_*(t) \coloneqq \int_\Omega \partial_t  c_N(\x, t) \dd \v{x}
  = \int_\Omega p  \, c_{N-1}(\x, t) \, \dd\v{x} \,,
\end{equation}
and the activation probability 
\begin{equation}
\label{eqn:prob}
  P_{*} \coloneqq \int_0^{\infty}  J_*(t)\, \dd t \,,
\end{equation}
from which the conditional activation flux $J_{*,\rm c}(t) = \Jbind(t)
/ \Pbind \,$ is derived \cite{chou_dorsogna_2014}.  Finally, the
conditional moments $ \mathds{E}[(\Tbind)^h]$ of the activation time
are
\begin{equation}\label{eqn:binding_time}
  \mathds{E}[(\Tbind)^h] := \int_0^\infty t^h \, J_{*,\rm c}(t) \,\dd t
  = \frac{\int_0^\infty t^h \, J_*(t) \, \dd t }{\int_0^\infty  J_* (t) \, \dd t}. \, \qquad h \geq 0 \,.
\end{equation}
The conditional mean activation time $\taubind$ and variance
$\sigmabind^2$ are obtained by substituting $h=1$ and $h=2$ into
\eqref{eqn:binding_time}, respectively,
\begin{equation}
\taubind =    \mathds{E}(\Tbind) \,, \qquad \sigmabind^2 =   \mathds{E}[[(\Tbind)^2]  - \taubind^2 \,.
\label{taubind}
\end{equation}
The time $\tau_{\rm nAPC}$ that a T cell spends engaged with a nAPC
before detaching and resuming its search for a cAPC can be estimated
by calculating the average time it takes a T cell starting in state
$\mathsf{N}_1$ to first reach the free state $\mathsf{T}_0$.  As
depicted in \eqref{eqn:transmission_pathway_APC}, the T cell performs
a random walk across the $N$ nAPC bound states detaching once
$\mathsf{T}_0$ is reached. In \cref{sec:param_transmission}, we show
the non-dimensional $\tau_{\rm nAPC}$ is
\begin{equation}
\label{MFPTtext}
\tau_{\rm nAPC} = \frac{1 - (p/q)^N}{1-p/q}.
\end{equation}
The interaction time in \eqref{MFPTtext} is an increasing function of
$N$, regardless of $p/q$, and an increasing function of $p/q$,
regardless of $N$. Thus, the more intermediate states there are along
the nAPC chain, the longer the T cell remains unproductively
engaged. Similarly, the larger the ratio of the forward to backward
rates $p/q$, the longer it will take for the T cell to return to the
free state.

\subsection{Neumann (perfectly reflecting) boundary conditions}
\label{sec:multistage_neumann}

We begin our analysis under perfectly reflecting, Neumann boundary
conditions by setting $\kappa = 0$ in \eqref{eqn:robin_bc_dim_nd}, and
calculate the activation probability $\Pbind$ in \eqref{eqn:prob}, and
the conditional mean activation time $\taubind$ in \eqref{taubind}.  For
an arbitrary function $y (\v{x}, t)$, $\v{x} \in \Omega$ we denote
$\overline{y}(t) = \int_\Omega y(\v{x},t) \, d\v{x}$.  By the
Divergence Theorem, we have
\begin{equation}
\label{Div}
\int_{\Omega} \Delta \rho_0 \, d\v{x} = \int_{\partial \Omega} \partial_n \rho_0 \, ds = 0 \,,
\end{equation}
where the left equality is due to the Neumann boundary condition for
$\rho_0$ in \eqref{eqn:robin_bc_dim_nd}.  Using \eqref{Div} and upon
integrating each equation in \eqref{eqn:multi-stage_nd} over the
entire domain $\Omega$, we obtain the following ODE system
\begin{equation}\label{eqn:kinetic_ODE_NBC_abb}
\begin{aligned}
  &\dfrac{\dd\overline{\v{y}}(t)}{\dd t} = \v{M} \, \overline{\v{y}}(t) \,, \quad
  \overline{\v{y}}(t) =
  \left(\overline{\rho}_0(t),\,
  \overline{\v{n}}(t), \, \overline{\v{c}}(t) \right)^T, \\[5pt]
  &\quad \overline{\v{y}}(0) = (1,0,\cdots,0) \in \mathbb{R}^{2N+1},
  \end{aligned}
\end{equation}
where $\overline{\v{n}}(t) \equiv (\overline{n}_{1}(t), \ldots,
\overline{n}_{N}(t))$ and $\overline{\v{c}}(t) \equiv
(\overline{c}_{1}(t), \ldots, \overline{c}_{N}(t))$.
The matrix $\v{M}$ is given by
\begin{equation}
\v{M} =\begin{bmatrix}
-\left[\mu_0 + (1 + f) k_{0,1} \right] & \v{e}_1^T & \v{e}_1^T \\[3pt]
k_{0,1}\,\v{e}_1 & \v{M}_\mathrm{n} & \textbf{O} \\[3pt]
f k_{0,1}\,\v{e}_1 & \textbf{O} & \v{M}_\mathrm{c}
\end{bmatrix} 
\end{equation}
where $\textbf{O}$ denotes the $N \times N$ zero matrix, and
$\v{M}_{\rm n}$ and $\v{M}_{\rm c}$ are the $N \times N$ tridiagonal
matrices defined in \eqref{eqn:multi-stage1} and
\eqref{eqn:multi-stage2}, respectively.  In this construction, $\v{M}$
is a $(2N + 1) \times (2N+1)$
matrix. Eq.~\eqref{eqn:kinetic_ODE_NBC_abb} can be solved by applying
the Laplace transform to Eq.~\eqref{eqn:kinetic_ODE_NBC_abb} with the
nonabsorbed states excluded (omitting $\overline{c}_{N}$ since it can
be determined from $\overline{c}_{N-1}$).
Upon defining the $N-1$ dimensional vector $\overline{\v{c}}'(s) =
(\bar{c}_1(s), \dots, \bar{c}_{N-1}(s))^{T}$ and taking the Laplace
transform of the truncated linear system
\eqref{eqn:kinetic_ODE_NBC_abb} that excludes $\bar{c}_{N}$, we find
\begin{equation}
\label{eqn:kinetic_ODE_NBC_laplace}
s \overline{\v{y}}'(s)  -\overline{\v{y}}'(t=0)= \v{M}' \, \overline{\v{y}}'(s) \,, \qquad \overline{\v{y}}'(s)
= \left(\overline{\rho}_0(s), \,\, \v{\overline{n}}(s),\,\,
\v{\overline{c}'}(s) \right)^T, 
\end{equation}
where $\v{M}'$ is the same as $\v{M}$ without the last row and last
column, and thus a $2N \times 2N$ matrix.
The formal solution to Eq.~\eqref{eqn:kinetic_ODE_NBC_laplace} is
\begin{equation}
\label{eqn:kinetic_ODE_NBC_laplace2}
\overline{\v{y}}'(s) = \big(s \v{\mathbb{I}} - \v{M}'\big)^{-1} \overline{\v{y}}'(t=0).
\end{equation}
Differentiating \eqref{eqn:kinetic_ODE_NBC_laplace} with respect to
$s$ and then setting $s=0$ gives
\begin{equation}\label{eqn:multi-stage_yprime}
\frac{\dd \overline{\v{y}}'(s)}{\dd s}\Big|_{s=0} = ({\v{M}'})^{-1} \, \overline{\v{y}}'(s=0).
\end{equation}
Eqs.~\eqref{eqn:kinetic_ODE_NBC_laplace2} and
\eqref{eqn:multi-stage_yprime} allow us to compute the activation
probability $P_*$ and the conditional mean activation time $\tau_*$,
both of which can be expressed in terms of the Laplace variable
through \eqref{eqn:binding_flux_multi-stage} to \eqref{taubind} as
\begin{eqnarray}
\label{eq:Pbindlap}
\Pbind &=& p \overline{c}_{N-1}(s=0), \\
\label{eq:taubindlap}
\taubind  &=& -\frac{1}{\overline{c}_{N-1}(s=0)}
\frac{\dd \overline{c}_{N-1}(s)}{\dd s}
      \Big|_{s=0}.
\end{eqnarray}
Upon setting $s=0$ in \eqref{eqn:kinetic_ODE_NBC_laplace2} and
inverting $\v{M}'$, we obtain $\overline{\v{y}}'(s=0)$; its last
entry, $\bar{c}_{N-1}(s=0)$, can be inserted into \eqref{eq:Pbindlap}
to obtain

\begin{equation}\label{eqn:multi-stage_Pbind}
  P_{*} = \frac{f k_{0,1} (p - q)}{fk_{0,1}(p - q)
    + \mu_0 \left[ p - q + 1 - \left(\displaystyle{\frac{q}{p}}\right) ^{N-1} \right]} \,.
\end{equation}
We similarly evaluate $\dd \overline{\v{y}}'(s) /\dd s$ for $s=0$ from
Eq.~\eqref{eqn:multi-stage_yprime}; its last entry and
$\overline{c}_{N-1}(s=0)$ are then substituted into
Eq.~\eqref{eq:taubindlap} to obtain $\taubind$.

\begin{figure}[htb]
\centering
\includegraphics[width=0.99\textwidth]{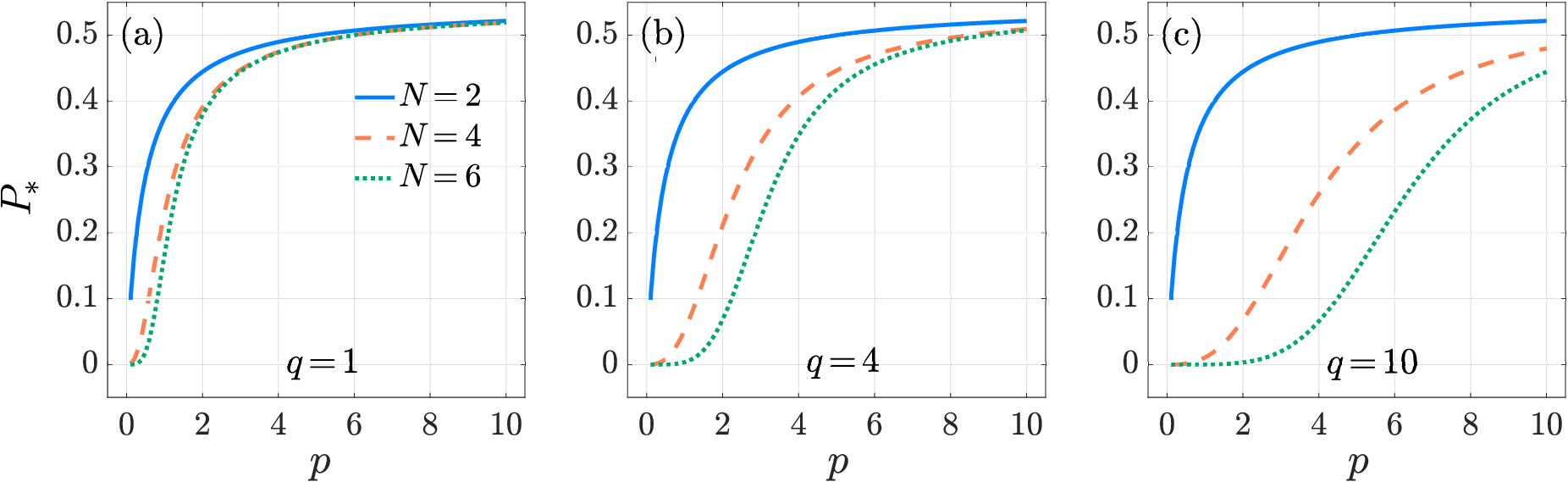}
\caption{\textbf{Multi-stage, Neumann boundary conditions}. The
  activation probability $P_{*}$ as a function of the forward binding
  rate $p$ in the multi-stage model \eqref{eqn:multi-stage_nd} for
  $N=2$ (blue-solid), $4$, (orange-dashed), and $6$ (green-dotted). We
  set $f = 5 \times 10^{-3}$, $k_{0,1} = 1$, $\mu_0 = 1/ 240$ and
  $q=1$ in panel (a), $q=4$ in panel (b) and $q=10$ in panel (c).  The
  values of $\Pbind$ are computed from \eqref{eqn:multi-stage_Pbind}
  All curves are monotonically increasing.}
\label{fig:multi-stage1}
\end{figure}

Results are shown in Fig.~\ref{fig:multi-stage1} for various choices
of $q$ and $N=2,4,6$. The activation likelihood $P_{*}$ is a
competition between degradation, expressed by $\mu_0$, and reaching
the final state of the cAPC arm. The schematic
\eqref{eqn:transmission_pathway_APC} reveals that for $N=2$ the
detachment rate $q$ affects the dynamics only as a T cell engages with
a nAPC. However, since the nAPC arm is characterized by reflecting
boundary conditions at its end state and no degradation is present, we
expect $P_{*}$ to be independent of $q$ for $N=2$.  This is observed
in Fig.~\ref{fig:multi-stage1}, where $P_{*}$ is also seen to be
increasing in $p$ for $N=2$.

Higher values of $p$ result in a higher likelihood that T cells
remains partially engaged to their cAPCs or to the other nAPCs, rather
than detaching from the first bound state, as can be seen by the form
$[\v{M}_{\v{n}}]_{ii} = [\v{M}_{\v{c}}]_{ii} = -(p +1)$.  In our
model, excursions along the nAPC arm do not increase the likelihood of
degradation, rather they shield T cells from the degradation that they
would experience in the unbound state $\msf{T}_0$.  At the same time,
higher $p$ values allow for T cells to be fully activated by their
cAPC by reaching state $\msf{C}_N$ in a more expedited manner,
escaping degradation while in the free state.  The overall effect of
increasing $p$ is thus to increase $\Pbind$. Since this picture
remains valid for other choices $N > 2$, the finding that $\Pbind$
increases with $p$ should be robust with respect to changes in $q$ and
$N$.  To understand how $\Pbind$ depends on $q$ (for $N>2$) and on
$N$, we note that as $q$ increases the likelihood of returning from
any intermediate state $\msf{N}_i$ or $\msf{C}_j$ with $i,j < N$ to
the unbound state $\msf{T}_0$, and thus for T cells to be degraded,
increases. Hence, $\Pbind$ should decrease with $q$ for fixed $p$ and
$N >2$. Similarly, we expect $\Pbind$ to decrease with increasing $N$
since the presence of more intervening steps to reach the activation
state $\msf{C}_N$ is associated with a greater likelihood for the T
cell to return to the free state $\msf{T}_0$ , where it is subject to
degradation.  The curves shown in Fig.~\ref{fig:multi-stage1} confirm
that $P_{*}$ is an increasing function of $p$ and a decreasing
function of $q$ and $N$ when all other quantities are kept
fixed. Finally, we verified that for all cases shown in
Fig.~\ref{fig:multi-stage1}, increasing the degradation $\mu_0$
decreases $\Pbind$ as expected.

\begin{figure}[htb]
\centering
\includegraphics[width=0.99\textwidth]{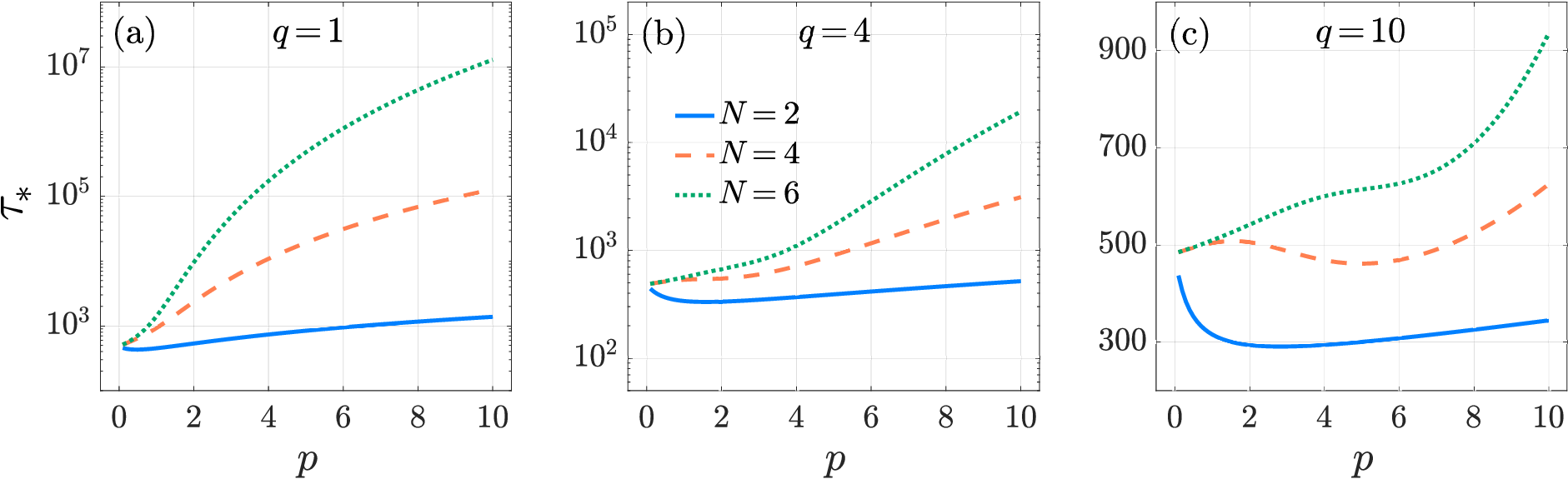}
\caption{\textbf{Multi-stage, Neumann boundary conditions}. The
  conditional mean activation time $\taubind$ as a function of the
  forward binding rate $p$ in the multi-stage model
  \eqref{eqn:multi-stage_nd} for $N=2$ (blue-solid) , $4$
  (orange-dashed), and $6$ (green-dotted). We set $f = 5 \times
  10^{-3}$, $k_{0,1} = 1$, $\mu_0 = 1/ 240$ and $q=1$ in panel (a),
  $q=4$ in panel (b) and $q=10$ in panel (c).  The values of
  $\taubind$ are computed from \eqref{eq:taubindlap} and
  \eqref{eqn:multi-stage_yprime}.  Depending on parameter choices,
  $\taubind$ can exhibit non-monotonic behavior.}
\label{fig:multi-stage2}
\end{figure}

In Fig.~\ref{fig:multi-stage2} we plot $\taubind$ as a function of $p$
for various values of $N$ and $q$.  We observe that $\taubind$
exhibits non-monotonic behavior as a function of $p$. This is because,
on one hand, increasing $p$ elongates the time a T cell remains on the
nAPC arm, increasing $\taubind$; on the other, it hastens the time for
the T cell to be fully activated at the end of the cAPC arm,
decreasing $\taubind$.  Which of these trends prevail depends on the
interplay between the magnitude of $p,q$ and the length $N$ of the
nAPC and cAPC arms.  This is illustrated by setting $N=2$, in which
case $\Pbind$ and $\taubind$ can be evaluated explicitly

\begin{eqnarray}
 \label{eq:N2P}
 \Pbind & = & \frac{f k_{0,1} p} { f k_{0,1} p + \mu_0 (p+1)},
 \qquad{\mbox {for $N=2$}}  \\[5pt]
\taubind & = & \frac {(1 + \mu_0 + p) q + 
 k_{0,1} (p+q) (p+1)  + f k_{0,1} q}
 {q \left[ f k_{0,1} p + \mu_0 (p+1) \right]},  \qquad{\mbox {for $N=2$}}.
 \label{eq:N2t}
\end{eqnarray}
Eqs.~\eqref{eq:N2P} and \eqref{eq:N2t} show that $\Pbind$
monotonically increases with $p$ but that $\taubind$, can be
non-monotonic in $p$ depending on the other parameters.  One
can evaluate the loci of the minima in $\taubind$ as a function of $q$
by taking the derivative of \eqref{eq:N2t} with respect to $p$.  We
find that, as a function of $p$, $\taubind$ displays a minimum only if
$q > q^*$ where
\begin{eqnarray}
  q^* = \frac{ f k_{0,1}^2 \mu_0 } { (f k_{0,1} + \mu_0)
    \left[f k_{0,1} (1 + k_{0,1} + f k_{0,1}) + 
2 f k_{0,1} \mu_0 +  \mu_0^2 \right]} < 1.
\end{eqnarray}
Non-monotonic behavior is observed in Fig.~\ref{fig:multi-stage2} for
$N>2$ as well.  Here, $\taubind$ is seen to increase with $p$, except
for intermediate $p$ values where a decreasing trend emerges. The
decreasing regimes correspond to optimal $p$ ranges where the T cell
can be activated by reaching the end of the cAPC arm while shortening
excursion times along the nAPC arm.  Although general analytical
estimates are not possible, we observe that the values of $p$ that
minimize $\taubind$ tend to increase with $N$.  The expression in
\eqref{eq:N2t} also reveals that for fixed $p$, $\taubind$ is a
decreasing function of $q$ for $N=2$.  Fig.\,\ref{fig:multi-stage2}
shows that $\taubind$ decreases with $q$ also for $N>2$.  Larger
values of $q$ diminish the time a T cell spends engaged with a nAPC
while having no effect on the time spent with its cAPC, so that larger
$q$ should lead to lower $\taubind$, as observed.
Fig.~\ref{fig:multi-stage2} also shows that $\taubind$ increases with
$N$.  In this case, increasing the length $N$ of the cAPC arm, results
in the T cell requiring a longer time to reach the final activation
stage.

Finally, we expect $\taubind$ to decrease as $f$ or $\mu_0$ is
increased.  Increasing $f$ increases the likelihood that the T cell
encounters its cAPC, shortening the time to full activation. Since
increasing $\mu_0$ hastens the degradation process, the conditional
mean activation time must be shorter to avoid degradation.  Both of
these trends in $\taubind$ are observed for all $p,q, N$ values
surveyed.

\subsection{Extreme first activation time statistics}

The results in section \ref{sec:multistage_neumann} for the activation
flux $J_*$, activation probability $P_*$, and conditional mean
activation time $\tau_*$ hold for a single T cell released at $\x=0$
at $t=0$. However, if the process is initialized with a collection of
$m$ T cells at the origin, one may wish to evaluate the probability
and first activation time of any T cell. Provided $m$ is not too large
as to significantly deplete the pool of free APCs, the T cells can be
considered independent particles.  For any specific T cell, we compute
the conditional survival probability $S_{\rm c}(t)$, defined as the
probability the T cell has not activated up to time $t$, given that it
will activate. We do so by explicitly solving $\mfrac{\dd S_{\rm
    c}(t)}{\dd t} = -J_{*, {\rm c}}(t)$ with the initial condition
$S_{\rm c}(0) = 1$ so that $S_{\rm c}(t) = 1- \int_{0}^{t}J_{*, \rm
  c}(t')\dd t'$.  From this quantity, we construct moments of the
first activation time of \textit{any} T cell.  First, we define the
probability that the minimum activation time $T_{\rm min}$ among all
$m$ initial T cells occurs after time $t$

  \begin{equation}
  \label{Pact1}
    \mathds{P}(T_{\rm min} > t)\coloneqq S_{\rm min}(t)=
     \big[(1-P_{*} ( 1 - S_{\rm c}(t)) \big]^{m} 
    \end{equation}
The term $P_{*} (1 - S_{\rm c} (t))$ represents the probability that a
T cell has activated, and has done so by time $t$.  $S_{\rm min}(t)$
is the overall survival function, the probability that none of the $m$
available T cells has been activated up to time $t$, regardless of
whether they will eventually be degraded, exit the T cell zone, or
activate.

We now subtract the probability that none of the
$m$ T cells activate from \eqref{Pact1} the probability that none of the
$m$ T cells activate, due to degradation or egress from the T cell
zone, given by $(1-P_{*})^{m}$.

\begin{figure}[t]
  \begin{center}
  \includegraphics[width=0.8\textwidth]{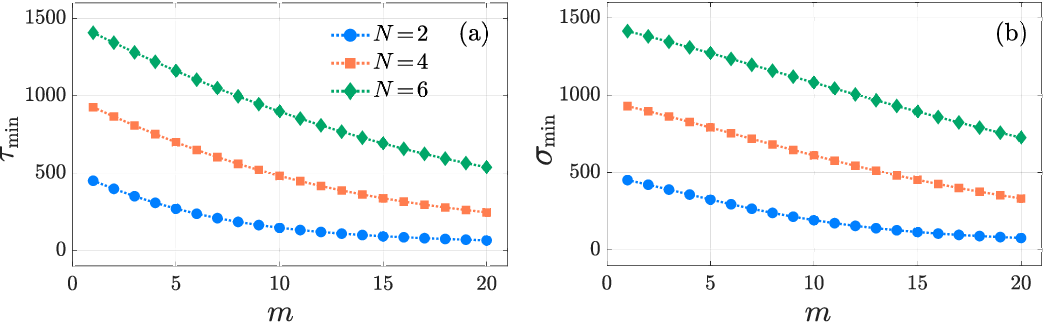}
\caption{\textbf{Multi-stage, Neumann boundary conditions}. The
  conditional mean and standard deviation of the minimum activation
  time of $m$ T cells in the multi-stage model
  \eqref{eqn:multi-stage_nd} for $N=2$ (blue-circle), $N=4$
  (orange-square), and $N=6$ (green-diamond).  We set $f = 5 \times
  10^{-3}$, $k_{0,1} = 1$, $\mu_0 = 1/ 240$, $p=q=1$.  The values of
  $\tau_{\rm min} $ and $\sigma_{\rm min}$ are computed from
  \eqref{first_moments} for $h=1$, and \eqref{sigmamin}, respectively.
  All curves are monotonically decreasing.}
  \label{fig:efpt}
\end{center}
\end{figure}

The difference is the probability that
$T_{\rm min}> t$ and at least one T cell activates.  This difference
is also the probability that $T_{\rm min} > t$ \textit{conditioned on}
at least one T cell activating, times the probability $P_{*}^{(m)}$
that at least one T cell activates. Since T cells are independent,
$P_{*}^{(m)} = 1-(1-P_{*})^{m}$ and

  \begin{equation}
    \begin{aligned}
\mathds{P}(T_{\rm min}<t\,\vert\, T_{\rm min} < \infty) \coloneqq & \,\, S_{\rm
  min}(t\vert T_{\rm min}<\infty) \\[5pt]
 = & \,\, \frac{\big[(1-P_{*} (1 -S_{\rm
      c}(t)) \big]^{m}-(1-P_{*})^{m}}{1-(1-P_{*})^{m}}.
    \end{aligned}
    \end{equation}
We now define the conditional first activation time distribution
$w_{\rm min}(t\vert T_{\rm min} < \infty) = -\frac {\dd S_{\rm min}(t\vert
T_{\rm min}<\infty)} {\dd t}$ and use it to compute moments of the first
activation time
\begin{equation}
    \mathds{E}\big[T_{\rm min}^{h}\big] = \int_{0}^{\infty} \!t^{h}
    w_{\rm min}(t\,\vert\, T_{\rm min}<\infty)\,\dd t, 
    \label{first_moments}
  \end{equation}
 and the corresponding standard deviation 
\begin{equation}
\label{sigmamin}
\sigma_{\rm min} := \sqrt{\mathds{E}\big[(T_{\rm min})^2\big] - \mathds{E}\big[T_{\rm min}\big]}. 
\end{equation}
In Fig.~\ref{fig:efpt} we plot the conditional mean $\tau_{\rm min} :=
\mathds{E}\big[T_{\rm min}\big]$ and the standard deviation
$\sigma_{\rm min}$. As expected, $\mathbb{E}\big[T_{\rm min}\big]$ decreases
with the number of searcher T cells $m$. The corresponding standard
deviation, $\sigma_{\rm min}$, also decreases with $m$ and remains
comparable in magnitude to the mean. This relationship suggests that a
Poisson distribution can reasonably approximate the conditional first
activation time distribution.

\subsection{Robin (partially reflecting) boundary conditions}
\label{sec:transmission_robin}

We now explore how $\Pbind$, $\taubind$ vary under Robin boundary
conditions, for finite $\kappa$ in \eqref{eqn:robin_bc_dim_nd}.  Since
in this case it is not possible to reduce \eqref{eqn:multi-stage_nd}
to a series of coupled ODEs, we must retain the inherent spatial
dependence.  Thus, we first write the formal time-dependent solution
to \eqref{eqn:multi-stage_PDE_nAPC_nd} and
\eqref{eqn:multi-stage_PDE_cAPC_nd}.  In Appendix
\ref{sec:eigendecomposition}, we show that the kinetic matrices
$\v{M}_\mathrm{n}$ and $\v{M}_\mathrm{c}$ are diagonalizable and can
be written as
\begin{equation}\label{eqn:kinetic_matrices}
  \v{M}_\mathrm{n} = \v{V}_\mathrm{n} \, \v{\Lambda}_\mathrm{n}
  \, \v{V}_\mathrm{n}^{-1} \,, \qquad \v{M}_\mathrm{c} = 
\v{V}_\mathrm{c} \, \v{\Lambda}_\mathrm{c} \, \v{V}_\mathrm{c}^{-1}, 
\end{equation}
where $\v{V}_\mathrm{n}$ and $\v{V}_\mathrm{c}$ are $N \times N$
matrices whose columns consist of the eigenvectors of
$\v{M}_\mathrm{n}$ and $\v{M}_\mathrm{c}$, respectively.  The diagonal
matrices $\v{\Lambda}_\mathrm{n} = \mathsf{diag}(\lambda_\mathrm{n}^1,
\lambda_\mathrm{n}^2, \dots, \lambda_\mathrm{n}^N)$ and
$\v{\Lambda}_\mathrm{c} = \mathsf{diag}(\lambda_\mathrm{c}^1,
\lambda_\mathrm{c}^2, \dots, \lambda_\mathrm{c}^N )$ consist of the
corresponding eigenvalues of $\v{M}_\mathrm{n}$ and $\v{M}_\mathrm{c}$
respectively.  Using the decompositions in
\eqref{eqn:kinetic_matrices} the solution components of
\eqref{eqn:multi-stage_PDE_nAPC_nd} and \eqref{eqn:multi-stage_PDE_cAPC_nd}
can be written as
\begin{equation}
  \begin{aligned}
n_i(\v{x}, t) & = k_{0,1} \, \sum\limits_{j=1}^N \,
      [\v{V}_\mathrm{n}]_{i,j} \, [\v{V}_\mathrm{n}^{-1}]_{j,1} \,
      \int_0^t e^{\lambda_\mathrm{n}^j (t-t')} \, \rho_0(\v{x},t') \, \dd t'
      \,, \quad 1 \leq i \leq N \,, \\[5pt]
c_i(\v{x}, t) & = f k_{0,1} \, \sum\limits_{j=1}^N \,
      [\v{V}_\mathrm{c}]_{i,j} \, [\v{V}_\mathrm{c}^{-1}]_{j,1} \,
      \int_0^t e^{\lambda_\mathrm{c}^j (t-t')} \, \rho_0(\v{x},t') \,
      \dd t' \,, \quad 1 \leq i \leq N \,.
  \end{aligned}
  \label{eqn:kinetic_ODE_sol2}
\end{equation}
By substituting the explicit expressions of $n_1$ and $c_1$ from
\eqref{eqn:kinetic_ODE_sol2} into \eqref{eqn:multi-stage_PDE_rho0_nd},
we obtain the integro-differential equation (IDE)
\begin{equation}\label{eqn:IDE}
\partial_t \rho_0 = D \Delta \rho_0
 - \left[\mu_0 + (1 + f) k_{0,1} \right]  \rho_0 +  \int_0^t {\cal K} (t-t') \, \rho_0(\v{x}, t') \, \dd t' \,,
\end{equation}
where the memory kernel $K(t)$ is defined by
\begin{equation}\label{eqn:kernel_transmission}
  {\cal{K}}(t)  = k_{0,1} \sum\limits_{j=1}^N [\v{V}_\mathrm{n}]_{1,j} \,
  [\v{V}_\mathrm{n}^{-1}]_{j,1} \, e^{\lambda_\mathrm{n}^j t}
  + f k_{0,1} \sum\limits_{j=1}^N [\v{V}_\mathrm{c}]_{1,j} \,
  [\v{V}_\mathrm{c}^{-1}]_{j,1} \, e^{\lambda_\mathrm{c}^j t} \,.
\end{equation}
We now use separation of variables as illustrated in \cref{sec:series}
to derive $\rho_0(\v{x}, t)$ under the Robin boundary condition in
\eqref{eqn:robin_bc_dim_nd}.  This quantity can then be used to
determine $n_i(\v{x}, t)$ and $c_i(\v{x}, t)$ for $i=1,2,\dots,N$ from
\eqref{eqn:kinetic_ODE_sol2}, finally calculating $\Pbind$ and
$\taubind$ via \eqref{eqn:binding_flux_multi-stage} to
\eqref{taubind}.

\begin{figure}[t]
\centering
\includegraphics[width=0.98\textwidth]{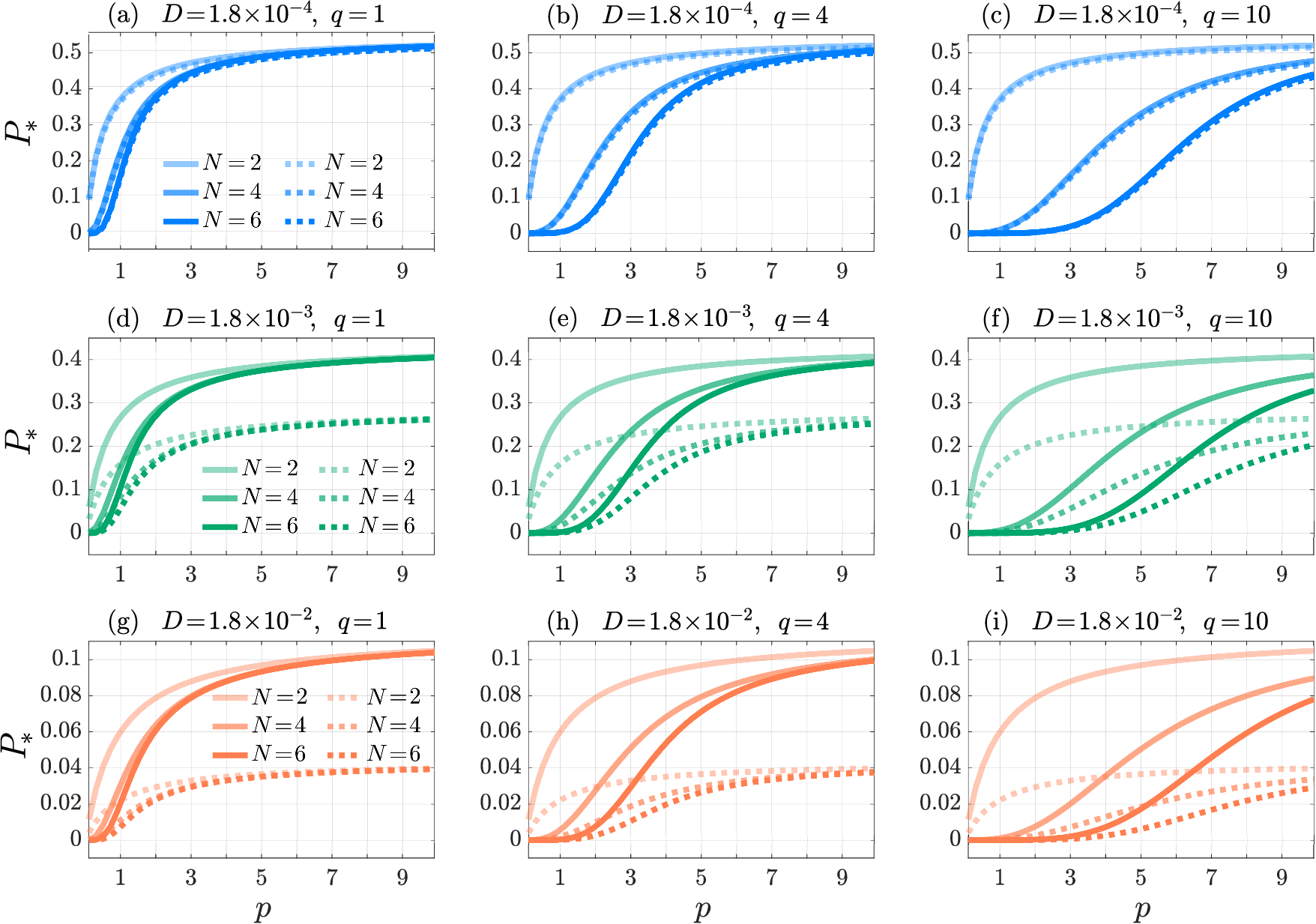}
\caption{\textbf{Multi-stage, Robin boundary conditions}. The
  activation probability $\Pbind$ as a function of the forward binding
  rate $p$ in the multi-stage model \eqref{eqn:multi-stage_nd} for
  $N=2,4,6$ (from lighter to darker shades) and Robin coefficient
  $\kappa =1$ (solid) and $\kappa \to \infty$ (dotted). We set $f = 5
  \times 10^{-3}$, $k_{0,1} = 1$, $\mu_0 = 1/ 240$, and $D = 1.8
  \times 10^{-4}$ (top row, blue), $1.8 \times 10^{-3}$ (middle row,
  green), $1.8 \times 10^{-2}$ (bottom row, orange), and $q=1, 4, 10$
  (from left to right).  The values of $\Pbind$ are computed from
  \eqref{eqn:prob}, \eqref{eqn:binding_flux_multi-stage},
  \eqref{eqn:kinetic_ODE_sol2}. All curves are monotonically
  increasing.  }
\label{fig:multi-stage_robin}
\end{figure}

Results are shown in Fig.~\ref{fig:multi-stage_robin} where the
activation probability $P_{*}$ is plotted as a function of $p$ for
various values of $q, N, \kappa, D$.  Trends that were observed under
the perfectly reflecting, Neumann boundary conditions in
Fig.~\ref{fig:multi-stage1} and that correspond to $\kappa=0$, are the
same. For example, $P_{*}$ remains an increasing function of $p$ and a
decreasing function of $q$ and $N$ when all other quantities are kept
fixed, mirroring the results in Fig.~\ref{fig:multi-stage1}.  The main
difference is that under Robin boundary conditions there is an extra
dependence on $\kappa$ and $D$.  First, we note that increasing
$\kappa$ should decrease the activation probability $P_{*}$ since a
larger $\kappa$ implies a higher likelihood that the T cell leaves the T cell zone before activation by its cAPC. We expect that
increasing the diffusion constant $D$ should also decrease $P_{*}$
since larger $D$ is equivalent to favoring transport over binding to
any APC.  This, in turn, would imply that the T cell is more likely to
be exposed to degradation, since the latter is only experienced in the
free form and not when the T cell is bound to any APC.  That $P_{*}$
should decrease with $D$ can be also be easily verified from the
explicit solution for $P_{*}$ in \eqref{finalrho0} of
\cref{sec:series}.  Fig.~\ref{fig:multi-stage_robin} confirms that
$P_{*}$ decreases as $\kappa$ and $D$ increase. As in
Fig.~\ref{fig:multi-stage1}, increasing the degradation rate $\mu_0$
decreases $\Pbind$ under Robin boundary conditions as well.

\begin{figure}[t]
\centering
\includegraphics[width=0.98\textwidth]{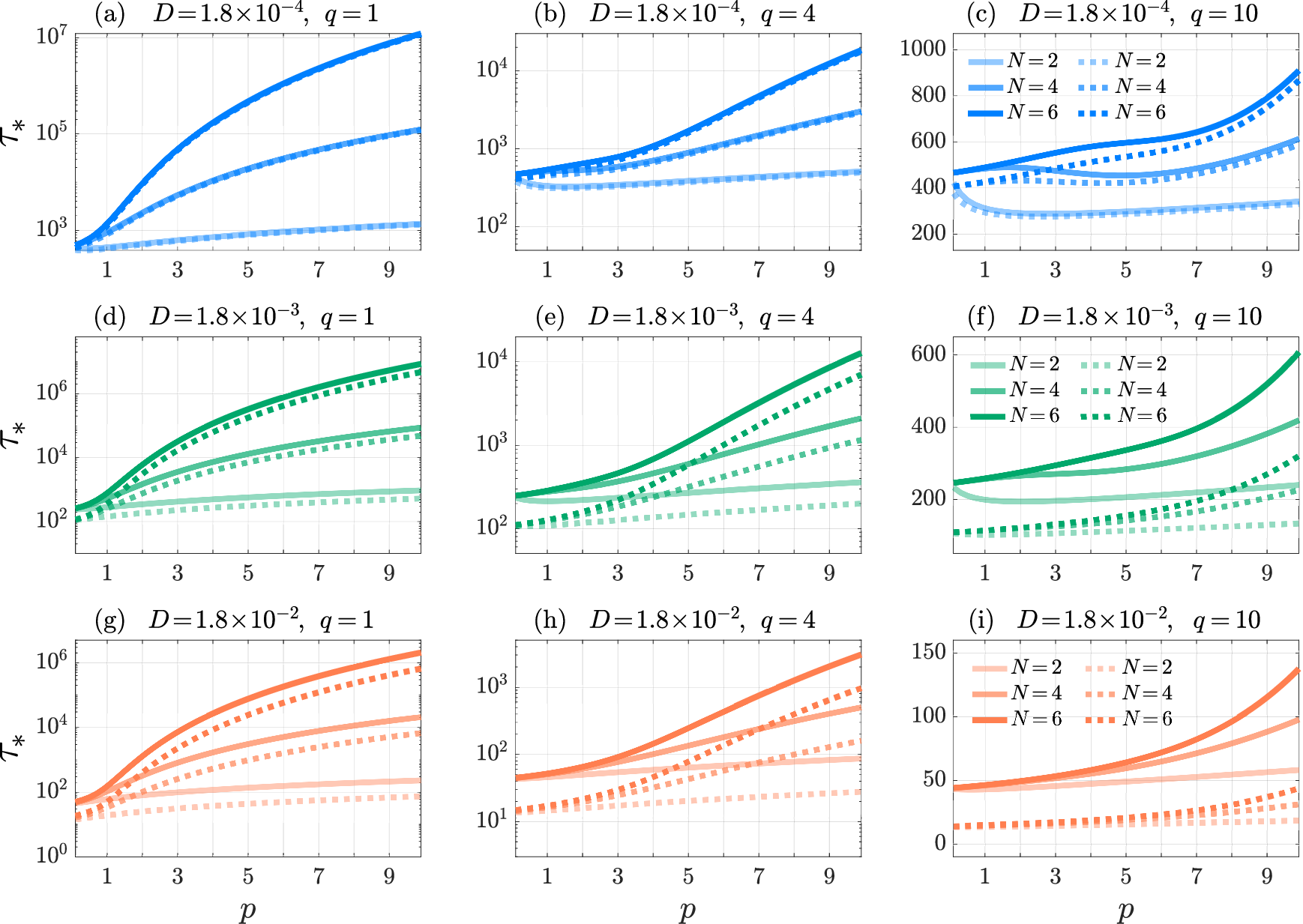}
\caption{\textbf{Multi-stage, Robin boundary conditions}. The
  conditional mean activation time $\taubind$ as a function of the
  forward binding rate $p$ in the multi-stage model
  \eqref{eqn:multi-stage_nd} for $N=2,4,6$ (from lighter to darker
  shades) and Robin coefficient $\kappa =1$ (solid) and $\kappa \to
  \infty$ (dotted). We set $f = 5 \times 10^{-3}$, $k_{0,1} = 1$,
  $\mu_0 = 1/ 240$, and $D = 1.8 \times 10^{-4}$ (top row, blue), $1.8
  \times 10^{-3}$ (middle row, green), $1.8 \times 10^{-2}$ (bottom
  row, orange), and $q=1, 4, 10$ (from left to right). The values of
  $\taubind$ are computed from \eqref{eqn:binding_time} for $h=1$,
  \eqref{eqn:binding_flux_multi-stage}, \eqref{eqn:kinetic_ODE_sol2}.
  Depending on parameter choices, $\taubind$ can exhibit non-monotonic
  behavior.  }
\label{fig:multi-stage3}
\end{figure}

Since increasing $\kappa$ or $D$ leads to faster escape, we expect the
conditional mean activation time $\taubind$ to decrease with $\kappa$
and $D$.  This dependence is shown in Fig.~\ref{fig:multi-stage3}
which plots $\taubind$ as a function of $p$ for various values of $q,
N, \kappa, D$.  As in Fig.~\ref{fig:multi-stage2}, we observe that
$\taubind$ remains a non-monotonic function of $p$, a decreasing
function of $q$, and an increasing function of $N$ when all other
quantities are kept fixed.

\section{Kinetic proofreading}
\label{sec:KP}

In the previous section, we assumed that the cAPC and nAPC multi-stage
arms are similar in that reactions proceed forward (with rate $p$) or
backward (with rate $q$) within both. At the end of the $N$-state
chain the T cell either binds irreversibly to its cAPC (activates), or
reaches an nAPC dead-end, as shown in
\eqref{eqn:transmission_pathway_APC}.  Here, we modify the previous
scheme in two ways.  First, we allow for T cells to completely
disengage from any of the intermediate steps along any APC chain to
return to the free state. Thus, when modeling interactions between T
cells and cAPCs, we replace the incremental backward step from
$\msf{C}_i$ to $\msf{C}_{i-1}$ with a return step from $\msf{C}_i$ to
$\msf{T}_0$. A similar modification is applied for steps along nAPC
arm.  Second, we assume that T cells can irreversibly bind to nAPCs at
the end of the nAPC arm as well, so that both end-states along the
cAPC and nAPC arms are absorbing. This is a canonical reaction scheme
that supports ``kinetic proofreading'' and is illustrated in
\eqref{KP_scheme_nAPC} and \eqref{KP_scheme_cAPC}.  Kinetic
proofreading (KPR) was first introduced in the 1970s to offer a
paradigm that could explain low error rates in DNA replication
\cite{hopfield_1974, ninio_1975}. Later, it was applied to study the
high specificity of T cell receptors in recognizing cAPCs
\cite{mckeithan_1995, lever_dushek_2014,li_chou_2024}.  Here, we
invoke the KPR mechanism to evaluate how the T cell enhances its
selection of the cAPC over nAPCs.

\begin{equation}
\label{KP_scheme_nAPC}
\begin{tikzcd}[row sep=3em, column sep=large]
\mathsf{T}_0 \arrow[r, "K_n"]
  & \mathsf{N}_1 \arrow[l, bend left, "{\small K_{1,0}}", pos=0.1]  
  \arrow[r, "P"]
  & \mathsf{N}_2 \arrow[r, "P"]
  \arrow[ll, bend left, "{\small K_{1,0}}", pos=0.2]
  & \cdots \arrow[r, "P"]
  & \mathsf{N}_{N-1} \arrow[r, "P"]
  \arrow[llll, bend left, "K_{1,0}", pos=0.2]
  & \mathsf{N}_N 
  \end{tikzcd}
  \end{equation}
\begin{equation}
\label{KP_scheme_cAPC}
\begin{tikzcd}[row sep=3em, column sep=large]
\mathsf{T}_0 \arrow[r, "f  K_n"]
  & \mathsf{C}_1 \arrow[l, bend left, "{\small \lambda K_{1,0}}", pos=0.1]  
  \arrow[r, "P"]
  & \mathsf{C}_2 \arrow[r, "P"]
  \arrow[ll, bend left, "{\small \lambda K_{1,0}}", pos=0.2]
  & \cdots \arrow[r, "P"]
  & \mathsf{C}_{N-1} \arrow[r, " P"]
  \arrow[llll, bend left, "\lambda K_{1,0}", pos=0.2]
  & \mathsf{C}_N 
  \end{tikzcd}
  \end{equation}


In dimensional units, the rate at which the T cell in any state
$\mathsf{N}_{i}$ disassembles and ``resets'' to the free state is
denoted $K_{1,0}$, along the nAPC arm.  The corresponding disassembly
or reseting rates are $\lambda K_{1,0}$ along the cAPC arm. We assume
that cAPC complexes are modestly more stable than nAPC complexes so
that their disassembly rates are slower, $\lambda < 1$. Because nAPCs
are more abundant than cAPCs ($f \ll 1$), T cells are kinetically more
likely to bind to nAPCs.  We utilize \eqref{nondim} to
nondimensionalize the return rates $K_{0,1}$ and $\lambda K_{0,1}$ to
$1$ and $\lambda$, respectively.  The non-dimensional kinetics of the
two-arm resetting model depicted in \eqref{KP_scheme_nAPC} and
\eqref{KP_scheme_cAPC} are described by
\begin{subequations}\label{eqn:PDE_KP_rho0_T}
\begin{eqnarray}
  \partial_t \rho_0(\x, t) & = & D \Delta \rho_0(\x,t) -
  \left[ \mu_0 + (1 + f) k_{0,1} \right] \rho_0 + \sum\limits_{i=1}^{N-1} n_i
  + \lambda \sum\limits_{i=1}^{N-1} c_i
\label{eqn:PDE_KP_rho0_0} \,, \\
\partial_t \, \v{n}(\x,t) & = & \v{B}_\mathrm{n} \v{n}
+ k_{0,1} \rho_0 \, \v{e}_1 \,, \label{eqn:PDE_KP_n} \\[5pt]
\partial_t \, \v{c}(\x,t) & = & \v{B}_\mathrm{c} \, \v{c}
+ f k_{0,1} \rho_0 \, \v{e}_1 \,. \label{eqn:PDE_KP_c}
\end{eqnarray}
\end{subequations}
The $N \times N$ bidiagonal matrices $\v{B}_{\rm n}$ and $\v{B}_{\rm
  c}$ describe interactions between T cells and APCs and incorporate
disassembly and adsorption at the last stage of the $N$-state chains. The
non-dimensional entries of $\v{B}_{\rm n}$ are

\begin{equation}
\label{eqn:KP_kinetic_matrices_n}
  \begin{aligned}
  &\begin{aligned}
[\v{B}_\mathrm{n}]_{i,i}  =
    \begin{cases}
      -(1+p) & \text{$1 \leq i \leq N-1$}, \\[5pt]
      0 & \text{$i=N$},
    \end{cases}
  \end{aligned}
  &\,\,\,  
  \begin{aligned}
    [\v{B}_\mathrm{n}]_{i,i-1} & = p \quad  2 \leq i \leq N-1,
  \end{aligned}
\end{aligned}
\end{equation}
whereas those pertaining to $\v{B}_{\rm c}$ are 
\begin{equation}
\label{eqn:KP_kinetic_matrices_c}
\begin{aligned}
  &\begin{aligned}
     [\v{B}_\mathrm{c}]_{i,i}  =
    \begin{cases}
      -(\lambda+p) & 1 \leq i \leq N-1, \\[5pt]
      0 & \text{$i=N$} \,,
    \end{cases}
  \end{aligned}
  &\,\,\,  
  \begin{aligned}
[\v{B}_\mathrm{c}]_{i,i-1} & = p \quad 2 \leq i \leq N-1.
  \end{aligned}
  \end{aligned}
\end{equation}
Finally, we employ the previous initial and boundary conditions,
Eqs.~\eqref{eqn:IC0} and \eqref{eqn:robin_bc_dim_nd}, respectively.
We now define $P_{*, {\rm n}}$ and $P_{*, {\rm c}}$ as the
probabilities that a T cell is ``improperly'' activated by an nAPC and
``properly'' activated by its cAPC, respectively.  To evaluate these
quantities, we first introduce the two fluxes into nAPC- and
cAPC-induced activation states

\begin{subequations}\label{eqn:KP_binding_fluxes0}
\begin{eqnarray}
J_{\rm n}(t) &:=& \int_{\Omega} \partial_t n_{N}(\v{x},t) d{\v{x}} =  \int_{\Omega} p \, n_{N-1}(\v{x},t) d{\v{x}} \,, \\[5pt]
J_{\rm c}(t) &:=& \int_{\Omega} \partial_t c_{N}(\v{x},t) d{\v{x}} =  \int_{\Omega} p  \, c_{N-1}(\v{x},t) d{\v{x}} \,.
\end{eqnarray}
\end{subequations}
The probabilities $\Pn$ and $\Pc$ are given by
\begin{equation}\label{eqn:KP_P}  
\Pn := \int_{0}^{\infty} J_{\rm n}(t) \dd t,  \qquad
\Pc := \int_{0}^{\infty} J_{\rm c}(t) \dd t \,.
\end{equation}
leading to the two conditional fluxes into the activation states
\begin{equation}
  \begin{aligned}
    J_{*, {\rm n}}(t)\coloneqq &  \displaystyle{\frac{J_{\rm n} (t) }{P_{*, {\rm n}}}} = \frac{1}{P_{*, {\rm n}}} \int_{\Omega} p \, n_{N-1}(\x, t) \dd {\v{x}}, \\
    J_{*,{\rm c}}(t)\coloneqq & \frac{J_{\rm c} (t) }{P_{*, {\rm c}}} =  \frac{1}{P_{*, {\rm c}}}\int_{\Omega} p  \, c_{N-1}(\x, t) \dd {\v{x}}.
\end{aligned}
    \label{eqn:KP_binding_fluxes}
  \end{equation}
The associated moments of the conditional first activation time to a
nAPC, $\mathds{E}\left[T_{*, {\rm n}}^h \right]$, and to a cAPC,
$\mathds{E}\left[T_{*, {\rm c}}^h \right]$, respectively, are
\begin{equation}\label{ETmc0}
\mathds{E}\left[T_{*,{\rm n}}^h \right]  =
\int_0^{\infty}\!\! t^h J_{*, {\rm n}}(t) \, \dd t, \quad 
\mathds{E}\left[T_{*, {\rm c}}^h \right] =
\int_0^{\infty} \!\!t^h J_{*,{\rm c}}(t) \, \dd t, 
\end{equation}
and the conditional mean activation times are obtained by setting $h=1$ in \eqref{ETmc0},

\begin{equation}\label{eqn:KP_tau}
\taun \coloneqq \mathds{E}\big[T_{*,\rm{n}} \big], \qquad
\tauc \coloneqq \mathds{E}\big[T_{*,\rm{c}} \big].
\end{equation}
Finally, we introduce the cAPC activation specificity 
$\Fc$, defined as the activation likelihood of a T cell to a cAPC
relative to the total activation likelihood
\begin{equation}
\label{eqn:KP_Fc_def}
\Fc = \frac{P_{*, {\rm c}}}{P_{*, {\rm n}} + P_{*, {\rm c}}} \,.
\end{equation}
Note that $\Fc =P_{*, {\rm c}}$ if Neumann boundary conditions are
applied and $\mu_0=0$ since in this case $P_{*, {\rm n}} + P_{*, {\rm
    c}} =1$. As $\Fc$ is the probability of activation by a cAPC,
given activation, we expect it to be independent of the degradation
rate $\mu_0$.

\subsection{Neumann (perfectly reflecting) boundary conditions}
\label{sec:neumannKPR}

\begin{figure}[htb]
\centering
\includegraphics[width=0.99\textwidth]{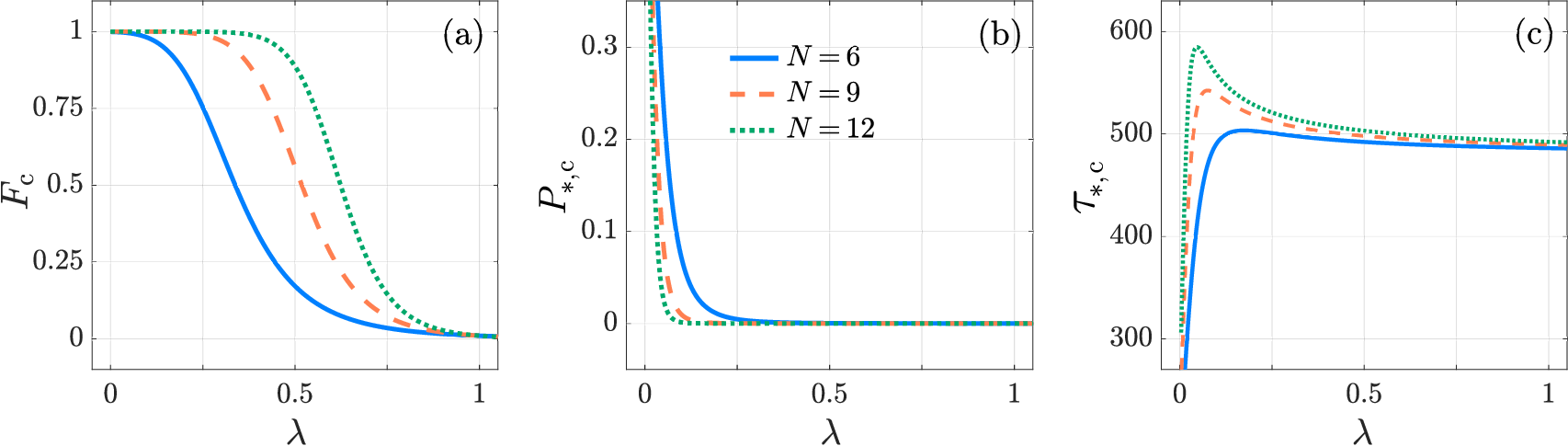} 
\caption{\textbf{Kinetic proofreading, Neumann boundary
    conditions}. The activation specificity $\Fc$, activation
  probability $P_{*, {\rm c}}$, conditional mean activation time
  $\tau_{*, {\rm c}}$, as functions of $\lambda$ for $N=6$
  (blue-solid), $9$ (orange-dashed) and $12$ (green-dotted) in the KPR
  model \eqref{eqn:PDE_KP_rho0_T}.  We set $f=10^{-2}$, $k_{0,1}=1$,
  $p=0.1$, $\mu_0 = 1/240$. While $\Fc$ and $P_{*, {\rm c}}$ are
  decreasing functions of $\lambda$, $\tau_{*, {\rm c}}$ shows
  non-monotonic behavior.  The largest value of $\Fc$ is at $\lambda
  =0$ and, for the given parameters, is $\Fc = f / [ f + (
    \frac{p}{1+p} )^{N-1} ] \approx 1$. The values of $\Fc$, $P_{*,
    {\rm c}}$, $\tau_{*, {\rm c}}$ are computed from
  \eqref{eqn:KP_Fc}, \eqref{eqn:KP_Pbind_c}, \eqref{eqn:KP_taubind_c},
  respectively.}
\label{fig:KP_neu_lambda}
\end{figure}

\begin{figure}[t]
\centering
\includegraphics[width=0.99\textwidth]{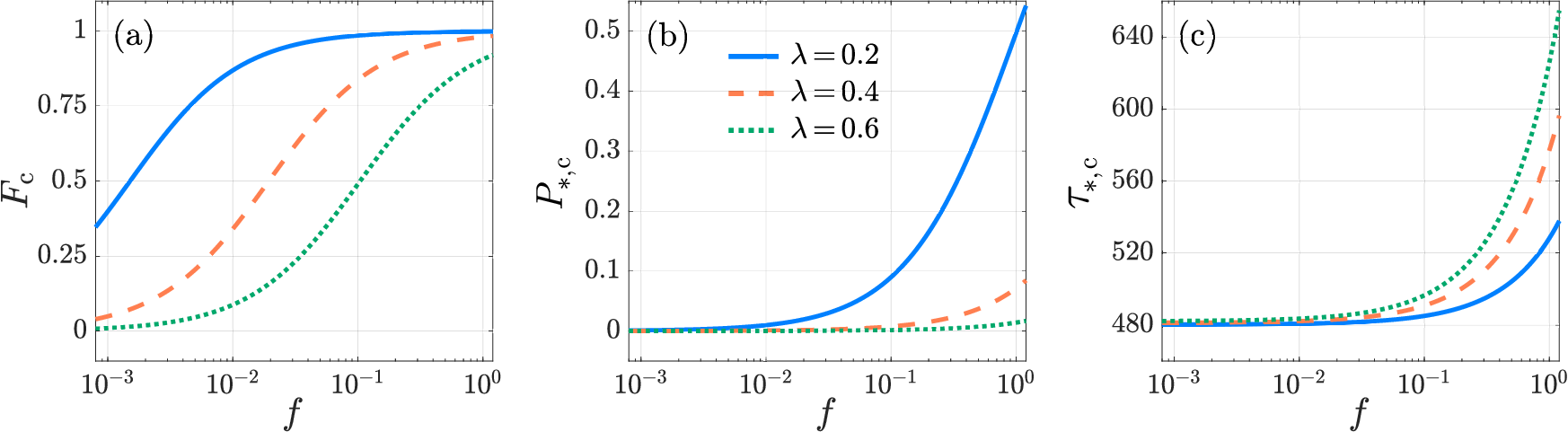} 
\caption{\textbf{Kinetic proofreading, Neumann boundary
    conditions}. The activation specificity $\Fc$, activation
  probability $P_{*, {\rm c}}$, and conditional mean activation time
  $\tau_{*, {\rm c}}$, as functions of $f$ for $\lambda=0.2$
  (blue-solid), $\lambda = 0.4$ (orange-dashed) and $\lambda = 0.6$
  (green-dotted) in the KPR model \eqref{eqn:PDE_KP_rho0_T}.  We set
  $N=6$, $k_{0,1}=1$, $p=0.1$, $\mu_0 = 1/240$.  The values of $\Fc$,
  $P_{*, {\rm c}}$, $\tau_{*, {\rm c}}$ are computed from
  \eqref{eqn:KP_Fc}, \eqref{eqn:KP_Pbind_c}, \eqref{eqn:KP_taubind_c},
  respectively.  All curves are monotonically increasing.}
\label{fig:KP_neu_f}
\end{figure}

\begin{figure}[htb]
\includegraphics[width=0.99\textwidth]{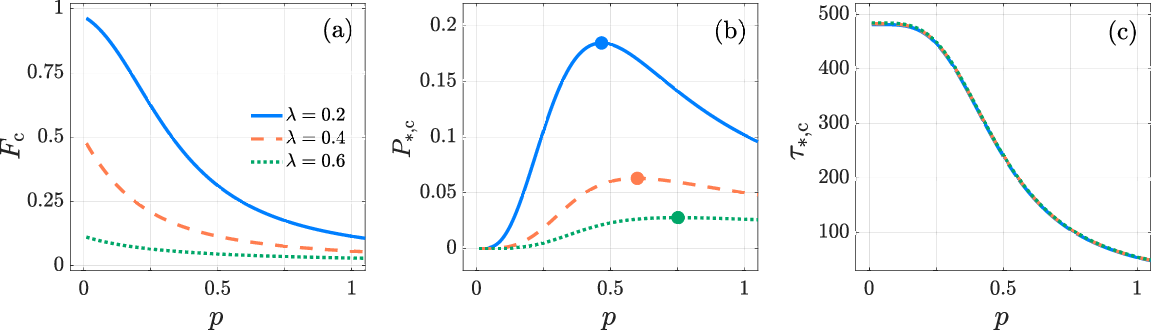}
\caption{\textbf{Kinetic proofreading, Neumann boundary
    conditions}. The activation specificity $\Fc$, activation
  probability $P_{*, {\rm c}}$, and conditional mean activation time
  $\tau_{*, {\rm c}}$ as functions of $p$ for $\lambda=0.2$
  (blue-solid), $0.4$ (orange-dashed) and $0.6$ (green-dotted) in the
  KPR model \eqref{eqn:PDE_KP_rho0_T}.  We set $N=6$, $f=10^{-2}$,
  $k_{0,1}=1$, $\mu_0 = 1/240$. In panel (a), $\Fc$ decreases with
  $p$; its largest value is at $p=0$ and is $\Fc = f / [f +
    \lambda^{N-1}]$.  The maximum of $P_{*, {\rm c}}$ in panel (b) is
  at $p = p_{\rm max}$ \eqref{eqn:KP_p_crit} and is represented by
  filled circles.  In panel (c), $\tau_{*, {\rm c}}$ exhibits similar
  trends as a function of $p$ for all three $\lambda$ values, leading
  to nearly indistinguishable curves.  The values of $\Fc$, $P_{*,
    {\rm c}}$, $\tau_{*, {\rm c}}$ are computed from
  \eqref{eqn:KP_Fc}, \eqref{eqn:KP_Pbind_c}, \eqref{eqn:KP_taubind_c},
  respectively.  }
\label{fig:KP_neu_p}
\end{figure}

\begin{figure}[htb]
\includegraphics[width=0.99\textwidth]{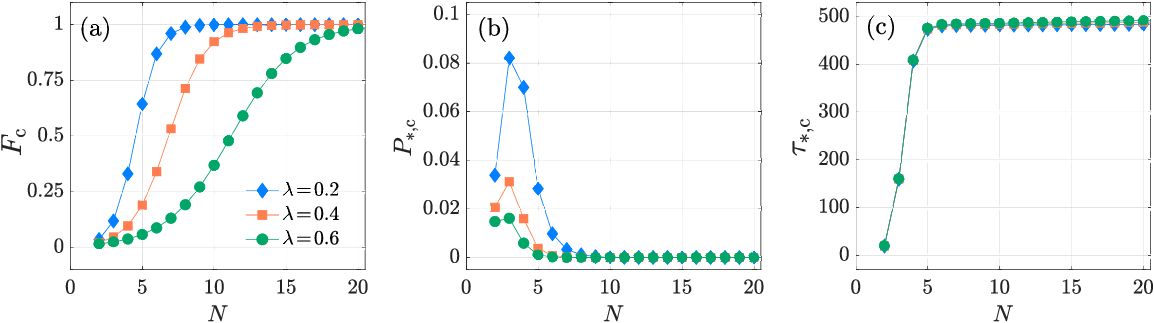}
\caption{\textbf{Kinetic proofreading, Neumann boundary
    condition}. The activation specificity $\Fc$, activation
  probability $P_{*, {\rm c}}$, and conditional mean activation time
  $\tau_{*, {\rm c}}$ as functions of $N \geq 2$ for $\lambda=0.2$
  (blue-diamond markers), $0.4$ (orange-square markers) and $0.6$
  (green-circular markers) in the KPR model
  \eqref{eqn:PDE_KP_rho0_T}. We set $f=10^{-2}$, $k_{0,1}=1$, $p=0.1$,
  $\mu_0 = 1/240$. In panel (a), $\Fc$ increases with $N$ and
  approaches its limiting value of one as $N \to \infty$. In panel
  (b), $P_{*, {\rm c}}$ attains its maximum at $N_{\rm max} = 3$ as
  per \eqref{eqn:KP_N_crit} for all three values of $\lambda$. In
  panel (c), $\tau_{*, {\rm c}}$ exhibits similar trends as a function
  of p for all three $\lambda$ values, leading to nearly
  indistinguishable curves.  The values of $\Fc$, $P_{*, {\rm c}}$,
  $\tau_{*, {\rm c}}$ are computed from \eqref{eqn:KP_Fc},
  \eqref{eqn:KP_Pbind_c}, \eqref{eqn:KP_taubind_c}, respectively. }
\label{fig:KP_neu_N}
\end{figure}

Predictions of the KPR mechanism under perfectly reflecting, Neumann
boundary conditions can be derived by setting $\kappa =0$ in
\eqref{eqn:robin_bc_dim}. The T cell can either be activated by its
cAPC, by an nAPC, or be degraded within the T cell zone.  Upon
integrating \eqref{eqn:PDE_KP_rho0_T} over $\Omega = B_1^3(0)$, akin
to the procedure used to derive \eqref{eqn:kinetic_ODE_NBC_abb}, we
obtain
\begin{equation}\label{eqn:kinetic_ODE_NBC_abbkpr}
  \dfrac{\dd \overline{\v{y}}(t)}{\dd t} = \v{B} \, \overline{\v{y}}(t), \,
  \quad \overline{\v{y}}(t) =
\left(\overline{\rho}_0(t),\, \overline{\v{n}}(t), \,\,
\overline{\v{c}}(t) \right)^T,
\end{equation}
where $\overline{\rho}_0, \overline{\v{n}} , \overline{\v{c}}$, and
their initial conditions are defined in
\eqref{eqn:kinetic_ODE_NBC_abb}.  The matrix $\v{B}$ in
\eqref{eqn:kinetic_ODE_NBC_abbkpr} is given by
\begin{equation}
\v{B} =\begin{bmatrix}
-\left[\mu_0 + (1 + f) k_{0,1} \right] &
(\v{1} - \v{e}_N)^T & \lambda(\v{1} - \v{e}_N)^T \\[3pt]
k_{0,1}\,\v{e}_1 & \v{B}_\mathrm{n} & \mathbf{O} \\[3pt]
f k_{0,1}\,\v{e}_1 & \mathbf{O} & \v{B}_\mathrm{c}
\end{bmatrix} 
\end{equation}
where $\textbf{O}$ is the $N \times N$ zero matrix, and $\v{B}_{\rm
  n}$ and $\v{B}_{\rm c}$ are the $N \times N$ matrices defined in
\eqref{eqn:KP_kinetic_matrices_n} and
\eqref{eqn:KP_kinetic_matrices_c}, respectively. The $N$-dimensional
vectors $\v{1}, \v{e}_1$ and $\v{e}_N$ are defined as $\v{1} =
(1,1,\cdots,1)^T$, $\v{e}_1 = (1,0,\cdots, 0)^T$ and $\v{e}_N =
(0,\cdots, 0, 1)^T \in \mathbb{R}^N$, respectively. We now eliminate
the equations for $\bar{n}_N$ and $\bar{c}_N$ in the ODE system
\eqref{eqn:kinetic_ODE_NBC_abbkpr}; upon taking the Laplace transform
of the truncated set of equations, we find
\begin{equation}\label{laplaceKP}
  s \overline{\v{y}}'(s) - \overline{\v{y}}'(t=0)
  = \v{B}' \v{y}'(s),\, \quad
  \overline{\v{y}}'(s) = \left(\overline{\rho}_0(s), \,
  \overline{\v{n}}'(s), \,\,
  \overline{\v{c}}'(s) \right)^T \,, 
\end{equation}
where $\v{B}'$ is the $(2N -1) \times (2N -1)$ matrix constructed by
eliminating the rows and columns of $\v{B}$ corresponding to
$\bar{n}_N$ and $\bar{c}_N$, respectively. These are the $(N+1)^{\rm
  th}$ and $(2N+1)^{\rm th}$ rows and columns of $\v{B}$.  Similarly,
$\overline{\v{n}}'(s)$ and $\overline{\v{c}}'(s)$ are the Laplace
transforms of $\overline{\v{n}}(t)$ and $\overline{\v{c}}(t)$ without
the $\bar{n}_N, \bar{c}_N$ entries.  Note, that since states $c_N$ and
$n_N$ are absorbing, the subsystem in \eqref{laplaceKP} is complete.
By evaluating the Laplace transforms we find
\begin{subequations}
  \label{eqn:KP_prob_neu}
\begin{eqnarray}
\label{eqn:KP_Pbind_n}
P_{*, {\rm n}} & = & \frac{k_{0,1}\left(\frac{p}{1+p}
  \right)^{N-1}}{\mu_0 + k_{0,1}\left(\frac{p}{1+p} \right)^{N-1}
  + fk_{0,1}\left(\frac{p}{\lambda+p} \right)^{N-1}} \,, \\[5pt]
P_{*, {\rm c}} & = & \frac{fk_{0,1}\left(\frac{p}{\lambda+p}
  \right)^{N-1}}{\mu_0 + k_{0,1}\left(\frac{p}{1+p} \right)^{N-1}
  + fk_{0,1}\left(\frac{p}{\lambda+p} \right)^{N-1}} \label{eqn:KP_Pbind_c} \,,
\end{eqnarray}
\end{subequations}
and
\begin{equation}
\tau_{*, {\rm n}} = \frac{N-1}{1 + p}  + T_*,  \quad
\tau_{*, {\rm c}} = \frac{N-1}{\lambda + p}  + T_*,
\label{eqn:KP_taubind_c}
\end{equation}
where
\begin{equation*}
T_* = \frac{1 + k_{0,1}\left[ 1 - \left(\frac{p}{1+p} \right)^{N-1}
    \left(1 + \frac{N-1}{1+p}\right) \right]
  + f k_{0,1}\left[ \frac{1}{\lambda}
    - \left(\frac{p}{\lambda+p} \right)^{N-1}
    \left(\frac{1}{\lambda}
    + \frac{N-1}{\lambda+p}\right) \right]}{\mu_0
  + k_{0,1}\left(\frac{p}{1+p} \right)^{N-1}
  + fk_{0,1}\left(\frac{p}{\lambda+p} \right)^{N-1}}.
\end{equation*}
The likelihood for a T cell to be degraded is $1 - P_{*, {\rm c}} -
P_{*, {\rm n}}$ and, as can be seen from \eqref{eqn:KP_Pbind_n} and
\eqref{eqn:KP_Pbind_c}, is zero if $\mu_0=0$.  Upon substituting
\eqref{eqn:KP_prob_neu} into \eqref{eqn:KP_Fc_def}, it can be verified
that the activation specificity is
\begin{equation}
\label{eqn:KP_Fc}
\Fc = \frac{f}{f + \left(\frac{\lambda+p}{1+p}\right)^{N-1}} \,.
\end{equation}
As expected, $\Fc$ does not depend on the degradation rate $\mu_0$.
After the initial binding, in the limit $\lambda \to 1$, the dynamics
along the two APC chains are equivalent and $\Fc \to f /(f +1) \ll 1 $
is the ratio of the initial binding rate of the T cell on the cAPC
arm, compared to the total initial binding rate on either APC.  $\Fc$
is a decreasing function of $\lambda$, which can be expected since
decreasing $\lambda$ hinders the ability of T cells to detach from the
cAPC chain, increasing the likelihood of activation by the
cAPC. Interestingly, for fixed $\lambda <1$, $\Fc$ is a decreasing
function of $p$; lower values of $p$ increase the time that a T cell
is bound to both APC chains, however, since $\lambda < 1$, detachment
is more likely along the nAPC chain than along the cAPC chain,
resulting in a higher likelihood of activation by the cAPC.  Note that
under the assumption $\lambda >1$, the opposite would hold, and $\Fc$
would be an increasing function of $p$.  Finally, as expected $\Fc$ is
an increasing function of $f$ and an increasing function of $N$ (if
$\lambda <1$). These trends are shown in Figs.~\ref{fig:KP_neu_lambda}
to \ref{fig:KP_neu_N}.  To achieve high specificity, {\textit {i.e.}}
to obtain $\Fc \to 1$, one must utilize small values of $\lambda$, $p$
and/or high values of $f$,$N$.

Although $\mu_0$ does not affect $\Fc$, it does reduce $P_{*, {\rm
    c}}$, as shown in \eqref{eqn:KP_Pbind_c}. Upon differentiating the
latter with respect to $p$ (or $N$) and keeping all other quantities
fixed, we find that $P_{*, {\rm c}}$ displays maxima at $p = p_{\rm
  max}$ and $N = N_{\rm max}$ as follows
\begin{eqnarray}
  p_{\rm max}  &=& \frac{1}{\left[\dfrac{k_{0,1}}{\mu_0}
      \left(\dfrac{1}{\lambda} -1 \right) \right]^{1/N} - 1 } \,, \qquad {\mbox {if}} \quad \lambda < \frac{k_{0,1}}{k_{0,1} + \mu_0} \label{eqn:KP_p_crit} \,, \\
\nonumber
\\[8pt]
N_{\rm max} &=& 1 + \left\lfloor\frac{\ln\left( \dfrac{k_{0,1}}{\mu_0}
  \cdot \dfrac{\ln(1 + p) -
    \ln(\lambda +p)}{\ln(\lambda + p) - \ln p} \right)}{\ln(1+p) - \ln p} \right\rfloor \,,  \label{eqn:KP_N_crit}
\\[5pt]
& {\mbox{if}}& \quad {\lambda <  - p + p^{\frac{\mu_0}{ k_{0,1} + \mu_0}}(1+p)^{\frac{k_{0,1}}{k_{0,1} + \mu_0}}} \,. \nonumber
\end{eqnarray}
Here, $\left\lfloor x \right\rfloor$ represents the floor function of
$x$. Similarly, it can be verified that $\Pc$ decreases with $\lambda,
\mu_0/k_{0,1}$ and increases with $f$.  The activation probability
$P_{*, {\rm c}}$ is plotted as a function of $\lambda, f, p, N$ in
Figs.~\ref{fig:KP_neu_lambda} to \ref{fig:KP_neu_N}.
Fig.~\ref{fig:KP_neu_lambda} reveals that the conditional mean
activation time $\tau_{*, {\rm c}}$ has a maximum in $\lambda$.  While
it is not feasible to determine analytical expressions for the value
of $\lambda$ that maximizes $\tau_{*, {\rm c}}$ from
\eqref{eqn:KP_taubind_c}, we note that for small $\lambda$, increasing
$\lambda$ is equivalent to a higher likelihood of T cells detaching
from the cAPC chain, extending the time required for activation.
However, as $\lambda$ continues to increase, competition from the nAPC
chain grows and T cells must bind to the cAPC more rapidly, leading to
decreasing values of $\tau_{*, {\rm c}}$.  Finally,
Figs.\,\ref{fig:KP_neu_f} to \ref{fig:KP_neu_N} show that $\tau_{*,
  {\rm c}}$ is a monotonic function of $f$, $N$ (increasing) and of
$p$ (decreasing).  We can also compare $\tau_{*, {\rm c}}$ and
$\tau_{*, {\rm n}}$ using \eqref{eqn:KP_taubind_c} to write
\begin{eqnarray}
\label{difftau}
 \tau_{*, {\rm c}} - \tau_{*, {\rm n}} =
 \frac{(1-\lambda)(N-1)}{(\lambda+p)(1+p)}.
 \end{eqnarray}
For $\lambda < 1$,  $\tau_{*, {\rm c}}$ is larger than $\tau_{*,
  {\rm n}}$ and vice-versa for $\lambda < 1$. Interestingly, $\tau_{*,
  {\rm c}} = \tau_{*, {\rm n}}$ for $\lambda =1$ regardless of the
value of $f$. Although the likelihood that a T cell initially binds
to a cAPC is lower than that of initially binding to a nAPC, once it
has bound,  if the return rates along both arms are equal ($\lambda =1$), 
then, on average,  the time it takes for a T cell to reach the final 
state of either the cAPC or nAPC arm, will be the same.  The
difference in \eqref{difftau} varies linearly with $N$ and is
appreciable only for large enough $N$. How $\Fc$,
$P_{*, {\rm n}}$, $P_{*, {\rm c}}$ depend on relevant parameters while
the others are kept fixed is summarized in Table \ref{tablefinal}.
The corresponding trends for $\tau_{*, {\rm n}}$ and $\tau_{*, {\rm
    c}}$ will depend on specific parameter choices when these
quantities are considered as functions of $\lambda, k_{0,1}$ or $f$; both
$\tau_{*, {\rm n}}$ and $\tau_{*, {\rm c}}$ will instead decrease with 
$\mu_0, p$ and increase with $N$, regardless of other parameters.

\begin{table}
    \resizebox{\columnwidth}{!}{
    \begin{tabular}{|c | c  | c| c| c| c| c| c}
    \hline  
  & {$\lambda$ } & {$p$ } &  $\mu_0$ & $k_{0,1}$  & {$f$ } & {$N$}\\[1ex]    
\hline
$\Fc$ 
 \eqref{eqn:KP_Fc} 
&  decrease &  decrease & $-$ & $-$ &  increase & increase \\[1ex]
\hline
$P_{*, {\rm c}}$   
\eqref{eqn:KP_Pbind_c} 
&  decrease & max at $p  = p_{\rm max}
 $ & decrease& increase &  increase & max at  $N = N_{\rm max}$\\[1ex]
\hline
$P_{*, {\rm n}}$   
\eqref{eqn:KP_Pbind_n} 
&  increase &  increase &  decrease & increase &  decrease
& decrease \\[1ex]
    \hline
  $\tau_{*, {\rm c}}$   
\eqref{eqn:KP_taubind_c} 
&    &  decrease &  decrease &   &  & increase \\[1ex]
   \hline
    $\tau_{*, {\rm n}}$   
\eqref{eqn:KP_taubind_c} 
&  & decrease &  decrease & 
& & increase  \\[1ex]
    \hline
    \end{tabular}}
    \caption{Trends of $\Fc$, $P_{*, {\rm c}}$, $P_{*, {\rm n}}$
      $\tau_{*, {\rm c}}$ and $\tau_{*, {\rm n}}$ as functions of
      $\lambda$, $p$, $\mu_0$, $k_{0,1}, f$, and $N$. Note that $\Fc$
      is independent of $\mu_0, k_{0,1}$ and that $P_{*, {\rm c}}$ and
      $P_{*, {\rm n}}$ depend only on the ratio $\mu_0/k_{0,1}$.
      Universal trends for $\tau_{*, {\rm c}}$ and $\tau_{*, {\rm n}}$
      can only be determined for $p, \mu_0, N$. Both quantities are
      decreasing functions of $p, \mu_0$ and increasing functions of
      $N$ if all other parameters are kept fixed. Whether
      $\tau_{*, {\rm c}}$ and $\tau_{*, {\rm n}}$ increase or decrease
      as functions of $\lambda, k_{0,1}, f$ depends on parameter
      choices.}
     \label{tablefinal}
    \end{table}

\subsection{Robin (partially reflecting) boundary conditions}
\label{Robin_KPR}

\begin{figure}[t]
\centering
\includegraphics[width=0.75\textwidth]{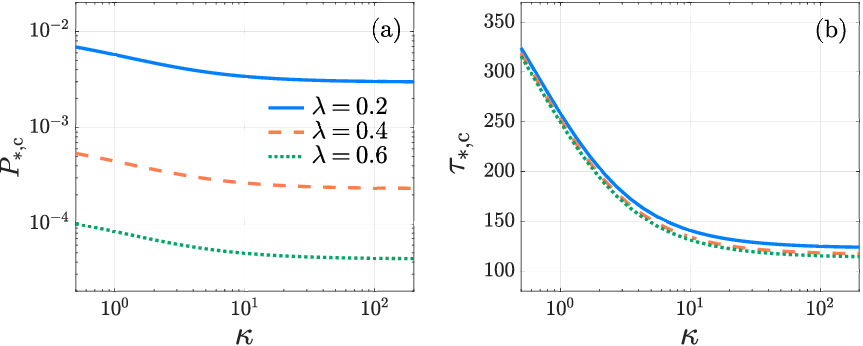} 
\caption{\textbf{Kinetic proofreading, Robin boundary conditions}. The
  activation probability $\Pc$ and the conditional mean activation
  time $\tauc$ as functions of the Robin coefficient $\kappa$ in the
  KPR model \eqref{eqn:PDE_KP_rho0_T} with $\lambda=0.2$ (blue-solid),
  0.4 (orange-dashed), and 0.6 (green-dotted). We set $N=6$, $\mu_0 =
  1/240$, $f=10^{-2}$, $k_{0,1}=1$, $D=1.8\times10^{-3}$.  The values
  of $P_{*, {\rm c}}$, $\tau_{*, {\rm c}}$ are computed from
  \eqref{eqn:KP_Pbind_c}, \eqref{eqn:KP_taubind_c}, respectively.  The
  activation specificity $\Fc$ in \eqref{eqn:KP_Fc} is independent of
  $\kappa$.  All curves are monotonically decreasing.}
\label{fig:KP_robin_lambda}
\end{figure}

We now analyze the KPR model under partially reflecting boundary
conditions, following the same approach used in section
\ref{sec:transmission_robin}. Specifically, we transform the PDE
system \eqref{eqn:PDE_KP_rho0_T} into an IDE for \(\rho_0 (\v{x}, t)\)
which, we will show, has the same form as \eqref{eqn:IDE} and where
the details of the KPR-dynamics are embedded in a new kernel
$\mathcal{K}_{\rm KP}(t)$.  We begin by noting that contrary to the
kinetic matrices $\v{M}_{\rm c}$ and $\v{M}_{\rm n}$ used in section
\ref{sec:transmission_robin}, the kinetic matrices $\v{B}_{\rm c}$ and
$\v{B}_{\rm n}$ in \eqref{eqn:KP_kinetic_matrices_n} are not
diagonalizable. We thus apply the Laplace transform to each equation
in \eqref{eqn:PDE_KP_n} and \eqref{eqn:PDE_KP_c} to obtain
\begin{subequations}
\begin{eqnarray}
  n_i(\v{x}, s) &=& \frac{k_{0,1} \, p^{i-1}}{(s + 1 + p)^i} \,
  \rho_0(\v{x},s) \,, \qquad \mbox{for} \enspace 1 \leq i \leq N-1 \,,
  \label{eqn:KP_n_expression}  \\[5pt]
c_i(\v{x}, s) &=& \frac{f k_{0,1} \, p^{i-1}}{(s + \lambda + p)^i} \, \rho_0(\v{x},s) \,, \qquad \mbox{for} \enspace 1 \leq i \leq N-1.  \label{eqn:KP_c_expression}
\end{eqnarray}
\end{subequations}
Upon adding all expressions for $n_i(\v{x}, s)$ in
\eqref{eqn:KP_n_expression} and $c_i(\v{x},s)$ in
\eqref{eqn:KP_c_expression} we find
\begin{equation}
  \begin{aligned}
\sum\limits_{i=1}^{N-1} n_i(\v{x},s)  =&
\frac{k_{0,1}}{s+1} \left[ 1 - \left(\frac{p}{s + 1 + p} \right)^{N-1}\right]
\rho_0(\v{x},s) \,, \\[10pt]
\sum\limits_{i=1}^{N-1} c_i(\v{x},s)  =&
\frac{f k_{0,1}}{s+\lambda} \left[ 1 - \left(\frac{p}{s + \lambda + p} \right)^{N-1}\right]
\rho_0(\v{x},s) \,.
  \end{aligned}
  \label{KP_conv} 
\end{equation}
The inverse Laplace transform \eqref{KP_conv} can be written as
\begin{equation}
  \begin{aligned}
  \sum\limits_{i=1}^{N-1} n_i(\v{x},t) = & \int_{0}^{t}\! \mathcal{K}_{\rm n}(t-t') \, \rho_0(\v{x},t') \, \dd t',\\
  \sum\limits_{i=1}^{N-1} c_i(\v{x},t) = & \int_{0}^{t}\! \mathcal{K}_{\rm c}(t-t') \, \rho_0(\v{x},t')  \, \dd t',
\end{aligned}
\label{KP_conv2}
\end{equation}
where the kernels $\mathcal{K}_{\rm n}(t)$ and $\mathcal{K}_{\rm
  c}(t)$ are the inverse Laplace transforms of
\begin{equation}
\mathcal{K}_{\rm n}(s) = \frac{k_{0,1}}{s+1} \left[ 1 -
  \left(\frac{p}{s + 1 + p} \right)^{N-1}\right], \,\,\, 
\mathcal{K}_{\rm c}(s)  = \frac{ \lambda f k_{0,1}}{s+\lambda} \left[ 1 - \left(\frac{p}{s + \lambda + p} \right)^{N-1}\right],
\label{kern1}
\end{equation}
respectively. Eqs.~\ref{kern1} define the KPR memory kernel
$\mathcal{K}_{\rm KPR}(t)$
\begin{equation}
\label{KP_kernel}
\mathcal{K}_{\rm KPR}(t) = \mathcal{K}_{\rm n}(t) + \mathcal{K}_{\rm c}(t) \,.
\end{equation}
Finally, by substituting \eqref{KP_conv2} into
\eqref{eqn:PDE_KP_rho0_T}, we re-obtain the IDE \eqref{eqn:IDE} with
the KPR memory kernel \eqref{KP_kernel}, $\mathcal{K}(t) \to
\mathcal{K}_{\rm KPR}(t)$.  Using the methods described in
\cref{sec:series} we derive the corresponding $\rho_0(\v{x},t)$, and
obtain $c_{N-1}(\v{x},t)$ using \eqref{eqn:KP_c_expression}.
Expressions~\eqref{eqn:KP_binding_fluxes0}--\eqref{eqn:KP_tau} allow
us to compute $P_{*, {\rm c}}$ and $\tau_{*,{\rm c}}$. Notably, the
resulting $\Fc$ is independent of $\kappa$ and still given by
\eqref{eqn:KP_Fc}.  To show this, we integrate
\eqref{eqn:KP_n_expression} and \eqref{eqn:KP_c_expression} spatially
over $\Omega$ for $i = N-1$, and evaluate the resulting expressions at
$s=0$ to find
\begin{equation}
\overline{n}_{N-1}(s=0) = \frac{k_{0,1} \,p^{N-2} \,\overline{\rho}_0(s=0)}{(1 + p)^{N-1}}, 
\quad \overline{c}_{N-1}(s=0) = \frac{f k_{0,1} \, p^{N-2}  \, \overline{\rho}_0(s=0)}{(\lambda + p)^{N-1}} \,. 
\end{equation}
Substituting these expressions into \eqref{eqn:KP_P} leads to
\begin{equation}
  \Pn = k_{0,1} \, \overline{\rho}_0(s=0)  \left(\frac{p}{1 + p} \right)^{N-1}\!\!\!\!\!\!\!, \quad\,\,
\Pc = f k_{0,1} \, \overline{\rho}_0(s=0) \left( \frac{p}{\lambda+ p} \right)^{N-1}\!\!\!\!\!\!\!\!\!.
\end{equation}
From the definition of $\Fc$ given in \eqref{eqn:KP_Fc_def} it follows
that \eqref{eqn:KP_Fc} still holds and that $\Fc$ is independent of
$\overline{\rho}_0(s=0)$ and of $\kappa$.  Similar to the results for
the multi-stage two-arm model in \cref{multi-stage}, $P_{*, \rm{n}}$,
$P_{*, \rm{c}}$, and $\tau_{*, \rm{c}}$ decrease as the Robin
coefficient $\kappa$ increases. Larger values of $\kappa$ imply a
higher likelihood that the T cell exits the T cell zone, resulting in
decreases in both $P_{*, \rm{n}}$ and $P_{*, \rm{c}}$.  Furthermore,
larger values of $\kappa$ require a T cell to reach its cAPC in a
shorter time, decreasing $\tauc$.  These trends are confirmed in
Fig.~\ref{fig:KP_robin_lambda} where $\Pc$ and $\tauc$ are observed to
both decrease with $\kappa$, for various values of $\lambda$. Although
$\Pc$ varies by about an order of magnitude across the different
choices of $\lambda$ in Fig.\,\ref{fig:KP_robin_lambda}, the
corresponding $\tauc$ are of similar scale.

\section{Discussion and Conclusions}
\label{discussion}

We constructed and analyzed a diffusion-reaction model to describe T
cells diffusing while seeking for their cognate APC target among a sea
of noncognate APCs. The search process is delayed by interactions with
nAPCs, and hindered by T cell death and escape from the T cell zone
compartment.  Our results show that when T cells and APCs bind through
a sequence of $N$ steps, the activation probability $P_*$ increases
with the forward-to-backward ratio $p/q$.  The bias toward forward
transitions along the APC chains decreases the likelihood of
degradation while in the free state, and increases the likelihood that
the final, activation state $N$ is reached.

How various parameters affect the conditional mean activation time
$\taubind$ is more subtle.  Within certain parameter regimes,
increasing $p$ may prolong interactions between T cells and nAPCs,
thereby increasing $\taubind$.  While $\Pbind$ consistently decreases
with the number of states $N$, $\taubind$ always increases with
$N$. As can be expected, a larger $N$ reduces the likelihood of
activation but extends the time required for it to occur.  Diffusion
does not directly promote activation but facilitates transport towards
the boundary, increasing the likelihood of escape if the boundary is
not fully reflecting.  As a result, both $\Pbind$ and $\taubind$
decrease with increasing $D$.

We also considered a variant T cell activation scheme in which
reaching the end of either the cAPC or nAPC interaction chain can
activate the T cell, but only in the first case in a successful
manner.  In this scheme, at each intermediate state along both the
nAPC and cAPC arms the T cell can reset back to the free state.  This
kinetic proofreading scheme biases the system toward successful cAPC
activation, even when the cAPC resetting rate is only modestly lower
than that of the nAPCs.

Mathematically, we studied an integro-differential equation with a
memory kernel derived from the multi-stage or KPR kinetics.  This IDE
reduces to an ODE under Neumann boundary conditions.  More general
kernels could be incorporated, using simplified or alternative
kinetics that account for differential lengths of the cAPC and nAPC
arms, heterogeneous forward and backward rates, and distinct nAPCs.
We also assumed that APC are uniformly distributed within the T cell
zone; this assumption could be modified to allow for heterogeneity in
the spatial distribution of APCs within the T cell zone.  Couplings
between lymph nodes and the vascular network could also be included,
to study T cells circulating through the lymphatic system.

\section{Declarations}

\subsection{Authorship and Contributionship}

All authors have read and approved the final manuscript.

\begin{itemize}
\setlength\itemsep{2pt}
\item \textit{Conceptualization}: M. R. D'Orsogna, T. Chou. 
\item \textit{Methodology}: T. Wong, I. Cho, M. R. D'Orsogna, T. Chou.
\item \textit{Software}: T. Wong, M. R. D'Orsogna.
\item \textit{Formal analysis and investigation}: T. Wong, M. R. D'Orsogna, T. Chou.
\item \textit{Writing - original draft}: T. Wong, I. Cho, M. R. D'Orsogna, T. Chou.
\item \textit{Writing - review and editing}: T. Wong, M. R. D'Orsogna, T. Chou.
\item \textit{Funding acquisition}: M. R. D'Orsogna, T. Chou.
\end{itemize}

\subsection{Conflicts of Interest}

The authors declare no conflicts of interest.

\subsection{Data and Code Availability}

No experimental data were used in this study. Codes are openly available at \url{https://github.com/kawahtony/t\_cell\_activation}.

\subsection{Ethics Statement on Informed Consent and Human/Animal Subjects}

This study did not involve human participants or animal subjects.

\subsection{Funding} 
This work was supported by the ARO through grant W911NF-23-1-0129
(MRD), the NSF through grant OAC-2320846 (MRD), the Simons Foundation
Institute through grant 815891 (TW), and the NIH through grant
R01HL146552 (TC).

\appendix

\section{Parameter estimation and physical considerations}
\label{parameters}
The size of a lymph node gland depends on where in the body it is
located; typically it is oval shaped.  Under healthy conditions, the
long axis ranges between 0.2 to 2.5 cm with an estimated average of
1.5 cm, whereas the typical short axis extends up to 1 cm
\cite{Magnusson1983, Dorfman1991, Ioachim1994, Steinkamp1995,
  Wendl2015, Okumus2017}.  The size of the T cell zone, where most of
the interactions between T cells and APCs occur, depends on the lymph
node anatomy, and whether it is activated or not, but in general the
it occupies a significant portion of the interior of a lymph node. In
this paper, we use the two terms interchangeably and a spherical
domain for the T cell zone. We set its radius to $a = 0.1$ cm. This
estimate is taken by assuming a typical short axis length of 0.4 cm
(corresponding to a radius of 0.2cm) and by assuming that the radius
of the T cell zone is roughly half that of short radius.

Na\"ive T cells measure between 5-10 $\mu$m in diameter, whereas the
size of APCs varies: B cells are in the same range as T cells, mature
dendritic cells have a diameter of 10-15 $\mu$m and for macrophages
the range is 20-50 $\mu$m.  A single lymph node contains between
10$^6$ to 10$^7$ T cells; the abundance increases upon activation when
the size of the T cells can also expand.  The number of APCs residing
in a lymph node depends on several factors, including the specific
type of antigen, the lymph node type, and whether an infection is
under way.

To estimate the dimensional parameters that appear in our model we
refer to experimental findings from previous literature. Several
imaging studies estimated the diffusivity of T cells to be about $D =
60 \, \mu \mathrm{m}^2 \mathrm{min}^{-1}$ \cite{miller_cahalan_2002,
  donovan_lythe_2012}.  On average it is estimated that T cells remain
in contact with nAPCs for about 3 minutes before dissociating
\cite{miller_cahalan_2002, Beltman_2007} so we set $K_{1,0} = 1/3 \,
\text{min}^{-1}$.  The mean residence time in the lymph node for a T
cell is between 12 and 24 hours \cite{Katakai2013, Miyasaka2016,
  Ugur2019, grigorova_cyster_2010}; thus, we set the exit rate from a
lymph node $M_0 = 1/720 \text{min}^{-1}$.  Finally, it is estimated
that during a 12 hour residence time in a lymph node, a T cell makes
about 160 interactions with nAPCs \cite{grigorova_cyster_2010}. Hence,
the average time separation between each T cell -- nAPC interaction is
about 4 and a half minutes.  Since we assume that the time a T cell
spends bound to a nAPC is 3 minutes, the time spent in the free
searching state is roughly one and a half minutes.

\section{Series solution for integro-differential equations}
\label{sec:series}
We solve the IDE \eqref{eqn:IDE} for a general kernel
${\cal{K}}(t)$. Due to spherical symmetry, we can omit the angular
variables and write $\rho_0({\bf x}, t) = \rho_0(r, t)$ so that
\eqref{eqn:IDE}, the boundary condition \eqref{eqn:robin_bc_dim_nd},
and the initial condition \eqref{eqn:IC0} become
\begin{subequations}\label{eqn:IDE_r}
\begin{align}
&\partial_t \rho_0 = \dfrac{D}{r^2} \partial_r \left(r^2 \partial_r \rho_0  \right) - \left[\mu_0 + (1 + f) k_{0,1} \right] \, \rho_0 + \int_0^{t} 
{\mathcal K} (t-t') \, \rho_0(r,t') \, \dd t' \,, \label{eqn:rho0_r} \\[5pt]
&\partial_{r} \rho_0 + \kappa \rho_0 = 0 \enspace \mbox{at } r = 1  \,, \qquad 
 \rho_0(r, 0) = \dfrac{\delta(r)}{4 \pi r^2} \,. \label{eqn:IC_BC}
\end{align}
\end{subequations}
To solve \eqref{eqn:rho0_r} we separate variables by setting
$\rho_0(r,t) = \phi(r) T(t)$ resulting in two decoupled problems
\begin{subequations}
\begin{eqnarray}
  &\frac{\dd^2 \phi}{\dd r^2} + \frac{2}{r} \frac{\dd\phi}{\dd r} = - \frac{
    \nu^2} {D} \phi \,, \label{eqn:phi_n} \\[5pt]
  &\frac{\dd T}{\dd t}
  + \left( \mu_0 + (1+ f ) k_{0,1} + \nu^2 \right) T = \int_0^T
  \mathcal{K}(t-s) T(s) \, \dd s \,, \label{eqn:T_n}
\end{eqnarray}
\end{subequations}
where $\nu$ is arbitrary. The solution to the spatial equation
\eqref{eqn:phi_n} is
\begin{equation}
\label{phileak}
\phi(r) = \frac 1 r \, 
\sin \left( {\frac {\nu r} {\sqrt{D}}} \right), 
\end{equation}
Applying this form to the boundary condition in \eqref{eqn:IC_BC} 
constrains $\nu$ to be any of the roots of the transcendental equation 
\begin{equation}
\label{boundarynondim}
(\kappa-1) \sin \left( \frac{\nu}{\sqrt D} \right) + \frac {\nu}{\sqrt
  D} \cos \left( \frac{\nu}{\sqrt D} \right) = 0, \qquad \nu >0.
\end{equation}
Eq.~\eqref{boundarynondim} can be rewritten as
\begin{equation}
\label{boundarynondim2}
\left( \kappa  - 1 \right) \sin \alpha + \alpha \cos \alpha = 0, \qquad  {\mbox {where \,}} \alpha =: \frac{\nu}{\sqrt{D}}, \quad \alpha >0. 
\end{equation}
We enumerate the $\alpha_n$ roots of \eqref{boundarynondim2} in
increasing order for $n \geq 1$. Once a specific root $\alpha_n$ is
selected, the related time dependent solution can be determined by
setting $\nu = \sqrt{D} \alpha_n$ in the temporal equation
\eqref{eqn:T_n}.  This leads to a series of solutions $ \phi_n(r)
T_n(t)$ associated to the specific root $\alpha = \alpha_n$.  To
proceed, and for simplicity, we set $N=1$ so that $n_1$ and $c_1$ can
be expressed in terms of $\rho_0$ through
\eqref{eqn:multi-stage_PDE_nAPC_nd} and
\eqref{eqn:multi-stage_PDE_cAPC_nd} as

\begin{subequations}
\begin{eqnarray}
  n_1(r,t) &=& \int_0^t k_{0,1} \, e^{-(t-t')} \, \rho_0(r, t') \, \dd t',
  \label{eqn:n1_sol} \\[5pt]
  c_1(r,t) &=& \int_0^t f k_{0,1} \, \rho_0(r, t') \, \dd t'.
  \label{eqn:c1_sol}
\end{eqnarray}
\end{subequations}
By substituting \eqref{eqn:n1_sol} into
\eqref{eqn:multi-stage_PDE_rho0_nd} the memory kernel ${\cal{K}} (t)$
can be explicitly written for $N=1$ as
${\cal{K}}(t) \equiv k_{0,1} \, e^{- t}$.
%
%
%
To find $T_n(t)$, we now take the Laplace transform $\mc{L}$ of
\eqref{eqn:T_n} for $N=1$ to find
\begin{equation}\label{eqn:T_n_laplace}
\mc{L}[T_n](s) = \frac{s+1}{(s + \mu_0 + (1 + f)  k_{0,1} + D \alpha_n^2) (s+1)  - k_{0,1}} 
\end{equation}
Where we used the fact that the Laplace transform of a convolution of
two functions is the product of the their Laplace transforms. The RHS
of \eqref{eqn:T_n_laplace} is a rational function of the Laplace
frequency variable $s$. Evaluating its inverse Laplace transform
yields
\begin{equation}\label{eqn:T_n_laplace2}
T_n(t) = \frac{1}{s_{n}^+ - s_n^-} \left[(1+s_n^+) e^{s_n^+t} - (1 + s_n^-) e^{s_n^- t} \right]
\end{equation}
\noindent
where $s_n^{\pm}$ are the (negative) roots of 
\begin{equation}\label{eqn:T_n_roots}
s^2 + ( \mu_0 + (1 + f)  k_{0,1} + D \alpha_n^2 + 1) s + (\mu_0+ f  k_{0,1} + D \alpha_n^2) = 0. 
\end{equation}
Upon rearranging terms we can write
\begin{equation}
\label{eqn:T_n_laplace3} 
T_n(t) = e ^{-h_n t} 
\left[ \frac{1 - h_n}{ H_n} \sinh(H_n t) + \cosh (H_n t)
\right]
\end{equation}
where $h_n = \mfrac{1}{2}(\mu_0 + (1 + f) k_{0,1} + D \alpha_n^2 +1)$
and $H_n = \mfrac{1}{2}\big[\mu_0 + (1 + f) k_{0,1} + D \alpha_n^2 +
  1)^2$ $- 4 (\mu_0+ f k_{0,1} + D \alpha_n^2)\big]^{1/2}$.
%
%
The solution in \eqref{eqn:T_n_laplace3} satisfies $T_n(t=0) =1$. The
overall solution can thus be written as a linear combination of
products $\phi_n(r) T_n(t)$ where the amplitudes $A_n$ are determined
by the spatial initial condition. We thus have
\begin{equation}
  \rho_0(r,t) = \sum\limits_{n=1}^\infty A_n \phi_n(r) T_n(t),\,\,\,
  \mbox{where}\,\,\, \sum \limits_{n=1}^\infty A_n \phi_n(r) = \frac{\delta(r)}{4 \pi r^2}.
\end{equation}
The $\phi_n(r)$ functions represent an orthogonal basis for $r \in
[0,1]$ since, given $n \neq m$
\begin{subequations}
\label{eqn:ortho}
\begin{eqnarray}
\qquad \qquad  \int_0^1 \phi_n(r)  \phi_m(r) \, 4 \pi r^2 \, \dd r &=& 
 4 \pi
  \int_0^1  
\sin \left(\alpha_n r \right)  \sin \left(\alpha_m r \right) \dd r\,,  \\[5pt] 
&=& 
4 \pi
\frac{\alpha_m
 \cos \alpha_m 
 \sin \alpha_n
 - \alpha_n  
 \cos \alpha_n   
 \sin \alpha_m}
{\alpha_n^2- \alpha_m^2} = 0 
\label{eqn:orthob}
\end{eqnarray}
\end{subequations} 
where the last equality in \eqref{eqn:orthob} follows from both
$\alpha_n, \alpha_m$ satisfying the boundary condition in
\eqref{boundarynondim2}. For $n \neq 0$
\begin{equation}
g_n^2 =:
 \int_0^1 \phi^2_n(r) 4 \pi r^2 \dd r =  
 4 \pi
 \int_0^1  \sin(\alpha_n r)^2 \dd r = 
\frac{  \pi \left[2 \alpha_n -  \sin(2 \alpha_n)\right]}
{\alpha_n}. 
\end{equation}
%
Since $\lim_{r \to 0} \phi_n(r) = \alpha_n$ we can write
\begin{equation}
A_n  =  \frac{ \alpha _n } {g_n^2} =  \frac {1}{\pi} \frac{\alpha_n^2} {\left[2 \alpha_n -  \sin(2 \alpha_n)\right]}
\quad {\mbox {for }} n \neq 0
\end{equation}
so that 
\begin{equation}
\label{finalrho0}
\rho_0(r,t) = \frac {1}{\pi} \sum\limits_{n=1}^\infty \frac{\alpha_n^2\, \sin(\alpha_n r )\,  e ^{-h_n t}}{ 
(2 \alpha_n -  \sin(2 \alpha_n))\,  r }
\left[ \frac{1 - h_n}{ H_n} \sinh(H_n t) + \cosh (H_n t) \right]  \,.
\end{equation}
It is straightforward to verify that $\lim_{t \to \infty}
\rho_{0}(r,t) = 0$ for all $r$ values.  The expression in
\eqref{finalrho0} can be inserted into Eqs.~\eqref{eqn:n1_sol},
\eqref{eqn:c1_sol} to find $n_1(r,t)$ and $c_1(r,t)$ respectively,
leading to the asymptotic limits $\lim_{t \to \infty} \rho_0(r,t) =
\lim_{t \to \infty} n_1(r,t) =0$ and
\begin{equation}
\lim_{t \to \infty} c_1(r,t) =  \frac {1}{\pi} \sum\limits_{n=1}^\infty \frac{ \alpha_n^2  \sin(\alpha_n r )\,}{ 
\big[2 \alpha_n -  \sin(2 \alpha_n)\big]\,  r }
\left[\frac{ f k_{0,1}}{\mu_0 + f k_{0,1} + D \alpha_n^2}
  \right] \label{finalspatial_c1a} 
 \end{equation}
We can calculate the activation probability $\Pbind$ under the Robin
boundary condition by evaluating the spatial integral in
\eqref{finalspatial_c1a}

\begin{eqnarray}
\label{Pbindk}
\Pbind &=&  \lim_{t \to \infty} \int_0^{1} c_1(r, t) \, 4 \pi r^2 \dd r. 
\end{eqnarray}
Upon inserting \eqref{finalspatial_c1a}  into \eqref{Pbindk}
and evaluating the integral we find 

\begin{eqnarray}
\Pbind &=& 4 \kappa  \sum\limits_{n=1}^\infty \frac{ \sin \alpha_n  }{ 2 \alpha_n -  \sin(2 \alpha_n) }
\left[\frac  { f k_{0,1}} {\mu_0 + f k_{0,1} + D \alpha_n^2} \right] \label{final_c1}
 \,.
\end{eqnarray}
As $\kappa \to 0$ one can show that $\alpha_ 1 \to 0$ as well and that
$\lim_{\kappa \to 0} \kappa \sin(\alpha_1) / (2 \alpha_1 - \sin(2
\alpha_1)) = 1/4$.  All other terms in \eqref{final_c1} converge to
zero as $\kappa \to 0$ so that \eqref{final_c1} reduces to
\begin{eqnarray}
\lim_{\kappa \to 0} \Pbind &=& 
\lim_{t \to \infty} \int_0^{1} c_1(r, t) \, 4 \pi r^2 \dd r
= \frac{ f k_{0,1}} {\mu_0 + f k_{0,1}}.
\label{final_c1k0}
\end{eqnarray}
Thus, under perfectly reflecting, Neumann boundary conditions, the
single-stage ($N=1$) activation probability is simply the ratio of the
binding rate between a T cell and its cAPC, to the total rate at which
the T cell reaches any of its absorbing states, either degradation or
binding to its cAPC.  The methods presented here are also applicable
to the general multi-stage case with $N >1$, where the memory kernel
${\cal K}(t)$ is given by \eqref{eqn:kernel_transmission}, and to the
KPR model, where the memory kernel ${\cal K}_{\rm KPR}(t)$ is given by
\eqref{KP_kernel}.

\section{Mean time of engagement between T cells and nAPCs}
\label{sec:param_transmission}
Here, we estimate the interaction time between a T cell and a nAPC by
determining the mean first time $\tau_{\rm{nAPC}}$ for a T cell at the
first stage of engagement $\mathsf{N}_1$ to return to the free state
$\mathsf{T}_0$.  For concreteness, we focus on the nAPC arm, as
extracted from \eqref{eqn:transmission_pathway_APC}

\begin{center}
\begin{equation}\label{eqn:transmission_pathway_nAPC}{
\schemestart
$\mathsf{T}_0$\arrow{<=>[\scriptsize $K_\mathrm{n}$][\scriptsize$K_{1,0}$]}$\mathsf{N}_1$\arrow{<=>[\scriptsize $P$][\scriptsize $Q$]}$\mathsf{N}_2$\arrow{<=>[\scriptsize $P$][\scriptsize $Q$]}$\cdots$\arrow{<=>[\scriptsize $P$][\scriptsize $Q$]} $\mathsf{N}_{N-1}$\arrow{<=>[\scriptsize $P$][\scriptsize $Q$]}$\mathsf{N}_N$
\schemestop}
\end{equation}
\end{center}
To determine $\tau_{\rm{nAPC}}$ we first write the equations for the
probability $\v{P} = (P_1, \dots, P_{N})^{T}$ for a T cell and a nAPC
to be bound at state $(1,\dots, N )$, respectively, and the
probability $P_0$ for the T cell to remain free, respectively. Since
we are not interested in transport phenomena \cite{hillen2025}, the
dynamics follow \eqref{eqn:multi-stage_PDE_nAPC} without the spatial
components. The forward equation is written as
\begin{equation}
\partial_t \, P_0 =  P_1 - (\mu_0 + k_{0,1})  P_0, \quad  
\partial_t \, \v{P} = \v{M}_\mathrm{n} \v{P} + k_{0,1} P_0 \, \v{e}_1\,.
\label{eqn:multi-stage_PDE2c}
\end{equation}
We assume that at $t=0$ the T cell is at state $n=1$ so that $P_1(t=0)
=1$ and $P_0(t=0)=P_{i \neq 1}(t=0) = 0$.  We also define the survival
probability $\v{S} = (S_1, \dots, S_{N})^{T}$ as the likelihood that
having started at state $(1,\dots, N )$ at $t=0$ the T cell is still
engaged to any nAPC state at time $t$.  Thus, the likelihood of the
free T cell being bound to the nAPC at any time $t$ is $S_0(t)=0$.
Furthermore if we assume the T cell is initially at its first binding
state then $S_1(t=0) =1$ and $S_{i \neq 1} (t=0) = 0$. It is well
known that the survival probability follows the backward equation
stemming from $\v{M}_{\mathrm n}^{\dagger}$, the adjoint of
$\v{M}_{\mathrm n}$ so that

\begin{equation} 
\partial_t \, S_0= 0, \qquad
\partial_t \, \v{S} = \v{M}_\mathrm{n}^{\dagger} \v{S}. 
\label{eqn:multi-stage_PDE_nAPCsurva} 
\end{equation}
The survival time distribution is given by $-\partial_t \,\v{S} $. 
As a result, the mean first passage time $\v{T} = (T_1, \dots, T_{N})^{T}$ 
of a partially bound T cell starting from stage $(1,\dots, N)$ to 
the free state can derived as
\begin{eqnarray}
\v{T} = - \int_0^{\infty} t \, \partial_t \v{S} \dd t =  \int_0^{\infty} \v{S} \, \dd t. 
\end{eqnarray}
Upon integrating the right-hand expression in
\eqref{eqn:multi-stage_PDE_nAPCsurva} with respect to time, from $t=0$
to $t \to \infty$, and by imposing $\v{S}(t =0) =1$ and $\v{S}(t \to
\infty) = 0$ we find $- \mathbb{I} = \v{M}_\mathrm{n}^{\dagger}
\v{T}$, where $\mathbb{I}$ is the identity matrix. The inverse $\v{T}
= - \left( \v{M}_\mathrm{n}^{\dagger}\right)^{-1} \mathbb{I}$ defines
the mean first passage time $T_1 \equiv \tau_{\rm nAPC}$ for the T
cell to be in the free, unbound state starting from the first bound
stage. Using standard matrix inversion methods we find
\begin{equation}\label{MFPTnAPC}
\tau_{\rm nAPC} = T_1 = \dfrac{1 - (p/q)^{N}}{1 - p/q}. 
\end{equation}
Note that $\lim_ {p/q \to 1} T_1 = N$.  As evaluated in
\eqref{MFPTnAPC}, $T_1$ is an increasing function of $p/q$, so that
the higher the bias away from the free state and towards higher
binding stages, the larger $T_1$.

\section{Spectral analysis of the kinetic matrices}\label{sec:eigendecomposition}
Here, we show that the kinetic matrices $\v{M}_{\rm n}$ and
$\v{M}_{\rm c}$ are diagonalizable and that they can be expressed
according to the eigendecomposition in \eqref{eqn:kinetic_matrices}.
One of the conditions for any $n \times n$ matrix to be diagonalizable
is that it must have $n$ distinct eigenvalues. For convenience, we
denote the $(N-1) \times (N-1)$ submatrix of $\v{M}_{\rm c}$ as
$\accentset{\sim}{\v{M}}_{\rm c}$ and denote its characteristic
polynomial as $\tilde{p}_{\rm c}(\lambda)$.  When $p,q > 0$,
$\v{M}_{\rm n}$ and $\accentset{\sim}{\v{M}}_{\rm c}$ are non-singular
Jacobi matrices, whose characteristic polynomials have distinct, real
and non-zero roots (see Chapter 2 of
Ref.~\cite{gantmacher_2002}). Particularly, $\v{M}_{\rm{n}}$ is
diagonalizable. A direct calculation shows that the characteristic
polynomial of $\v{M}_{\rm c}$, $p_{\rm c}(\lambda) = \lambda \,
\tilde{p}_{\rm c}(\lambda)$, has distinct, real and non-zero roots,
implying that it also has distinct, real eigenvalues, one of them
being zero. The matrix $\v{M}_{\rm c}$ is therefore diagonalizable. In
conclusion, the kinetic matrices $\v{M}_{\rm n}$ and $\v{M}_{\rm c}$
are diagonalizable provided that the forward and backward binding
rates $p,q$ are positive.

\section{Further reading} 

Background on T cell activation can be found in \cite{Parkin2001,
  Blattman2002, Jenkins2010, grigorova_cyster_2010,
  krummel_gerard_2016}. Experimental imaging studies of T cell–APC
interactions are reported in \cite{miller_parker_2003,
  miller_cahalan_2002, miller_parker_2004,
  stoll_germain_2002}. Mathematical and computational models of T cell
migration are presented in \cite{donovan_lythe_2012,
  preston_pritchard_2006, harris_hunter_2012}.  First-passage time
theory is introduced in
\cite{redner_2001,chou_dorsogna_2014,iyer_zilman_2016}. Kinetic
proofreading was proposed in \cite{hopfield_1974, ninio_1975} and
applied to antigen recognition in \cite{mckeithan_1995}.

\bibliographystyle{siamplain}
\bibliography{references}
\end{document}